\documentclass[a4paper,prl,superscriptaddress,floatfix,nofootinbib,twocolumn]{revtex4-1}

\usepackage{bbold}
\usepackage{bbm}
\usepackage[pdftex]{graphicx}
\usepackage{latexsym,amsmath,verbatim,amssymb,txfonts, mathtools}
\usepackage{color}
\usepackage{rotating}
\usepackage{verbatim}
\usepackage{multirow}
\usepackage[english]{babel}
\usepackage{comment}
\usepackage{hyperref}
\usepackage{mathrsfs}

\newcommand{\SM}{SM}
\newcommand{\av}[1]{\langle {#1} \rangle}

\newcommand{\change}[1]{{#1}}
\newcommand{\changejan}[1]{{#1}}

\begin{document}

\title{Percolation theory of self-exciting temporal processes}

\author{Daniele Notarmuzi}
\affiliation{Center for Complex Networks and Systems Research, Luddy School
  of Informatics, Computing, and Engineering, Indiana University, Bloomington,
  Indiana 47408, USA}

\author{Claudio Castellano}
\affiliation{Istituto dei Sistemi Complessi (ISC-CNR), Via dei Taurini 19, I-00185 Roma, Italy}

\author{Alessandro Flammini}
\affiliation{Center for Complex Networks and Systems Research, Luddy School
  of Informatics, Computing, and Engineering, Indiana University, Bloomington,
  Indiana 47408, USA}

\author{Dario Mazzilli}
\affiliation{Center for Complex Networks and Systems Research, Luddy School
  of Informatics, Computing, and Engineering, Indiana University, Bloomington,
  Indiana 47408, USA}
  
  \author{Filippo Radicchi}
\email{filiradi@indiana.edu}
  \affiliation{Center for Complex Networks and Systems Research, Luddy School
  of Informatics, Computing, and Engineering, Indiana University, Bloomington,
  Indiana 47408, USA}

\begin{abstract}
We investigate how the properties
 of inhomogeneous patterns of activity, appearing in
many natural and social phenomena, depend 
on the temporal resolution used to define individual bursts of activity.
To this end,  we consider
time series of microscopic events produced by a
self-exciting Hawkes process,
and leverage a percolation framework to study the formation of macroscopic bursts of activity as a function of the resolution parameter. We find that 
the very same process may result in 
different distributions of avalanche size
and duration, which are understood in terms of
the competition between the 1D percolation and the branching process
universality class. 
Pure regimes for the individual classes are observed at specific values of the resolution parameter corresponding to the critical points of the percolation diagram. A regime of crossover characterized by a mixture of the two universal behaviors is observed in a wide region of the diagram. The hybrid scaling appears to be a likely outcome for an analysis of the time series based on a reasonably chosen, but not precisely adjusted, value of the resolution parameter. 
\end{abstract}

\maketitle

%% Introduction

% Avalanches and power laws
Inhomogeneous patterns of activity, characterized by bursts of events separated by periods
of quiescence, are ubiquitous
in nature~\cite{karsai2018bursty}. 
The firing of neurons~\cite{dalla2019modeling, beggs2003neuronal},
earthquakes~\cite{bak2002unified}, 
energy release in astrophysical systems~\cite{wang2013self} and 
spreading of information in social 
systems~\cite{barabasi2005origin, karsai2012universal,
  gleeson2014competition} exhibit bursty activity, with
intensity and duration of bursts obeying power-law
distributions~\cite{dalla2019modeling, beggs2003neuronal, karsai2012universal}. 

If activity consists of point-like events in time, 
size and duration of bursts
are obtained from the inter-event time sequence. The analysis of many
systems~\cite{fontenele2019criticality,barabasi2005origin,karsai2012universal,
  weng2012competition,bak2002unified} reveals that
the inter-event time between consecutive events has a fat-tailed
distribution~\cite{barabasi2005origin, bak2002unified, karsai2012universal}.
This distribution appears more reliable for the characterization of
correlation in bursty systems than other traditional measures, e.g., the autocorrelation
function~\cite{karsai2012universal, jo2015correlated, kumar2020interevent}. 
However, the relation between autocorrelation and
burst size distribution is opaque. 
Further complications arise as the separation between different bursts
is not clear-cut. In discrete time series, avalanches of correlated
activity are monitored by coarsening the time series at a fixed
temporal scale, and correlations are measured by
assigning events to the same burst if their inter-event time is
smaller than a given threshold ~\cite{karsai2012universal}.
The threshold is set equal to some 
arbitrarily chosen
value and/or imposed by the temporal resolution at which empirical
data
are acquired, despite its potential of affecting the properties of
the resulting distributions\change{~\cite{pasquale2008self, neto2019unified, miller2019scale, chiappalone2005burst, levina2017subsampling, janicevic2016interevent, villegas2019time}.}
%~\cite{janicevic2016interevent, villegas2019time}. 

The purpose of the present letter is to
understand the relation between temporal resolution and burst statistics.
\change{We introduce a principled technique
to determine the value of the time
resolution that should be used to define avalanches from time
series. We validate the method on time series generated according to an}
Hawkes process~\cite{hawkes1971spectra}, a model 
of autocorrelated behavior used for the description of
earthquakes~\cite{helmstetter2002subcritical, ogata1988statistical},
neuronal networks~\cite{kossio2018growing},
and socio-economic systems~\cite{crane2008robust, sornette2004endogenous}. 
The use of the Hawkes process
affords us a complete control over the mechanism that generates
correlations
and the possibility to attack the problem analytically.   

We start by defining a cluster of activity consistently with
the informal notion of a burst composed of close-by events.
Data are represented by $K$ total events $\{t_1, \ldots, t_K\}$,
where $t_i$ is the time of appearance of the $i$-th event.
We fix a resolution parameter $\Delta \geq 0$ to identify clusters of
activity. A cluster starting at time $t_b$
is given by the $S$ consecutive events $\{t_{b}, t_{b+1}, \ldots, t_{b+S-1} \}$ 
such that $t_b - t_{b-1} > \Delta$, $t_{b+S} - t_{b+S-1} > \Delta$, and 
$t_{b+i} - t_{b+i-1} \leq \Delta$ for all $i=1, \ldots, S$. We assume 
$t_{0} = -\infty$ and  $t_{K+1} = + \infty$, implying that the first
and the last events
open and close a cluster, respectively. We define the size $S$ as the  number of events within the cluster, and its duration as $T = t_{b+S-1} - t_{b}$, i.e.,  the time lag between the first and last event in the cluster. 

If $\Delta$ is 
larger
than the largest inter-event time, then we have a single cluster of
size $K$ and duration $t_K-t_1$. On the other hand, if $\Delta$ is
smaller than the smallest inter-event time, each event is a cluster of size 1 and duration 0. 
As in 1D percolation problems~\cite{stauffer2018introduction}, we expect for an intermediate value $\Delta=\Delta^*$ a transition from the non-percolating to the percolating phase.
What can we learn from the percolation diagram of the time series?
Does fixing $\Delta = \Delta^*$ allow us to observe properties of the
process otherwise not apparent? 

%%%%%%%%%%%%%%

%%Hawkes process

\begin{figure*}[!htb]
\begin{center}
\includegraphics[width=0.85\textwidth]{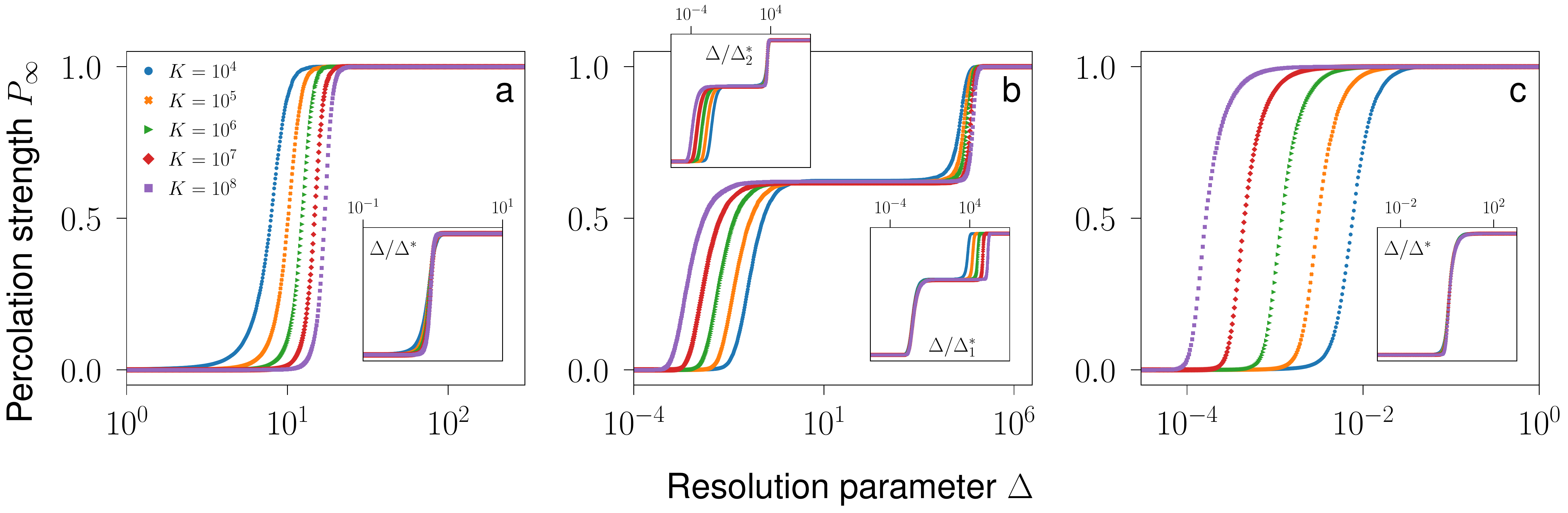}    
\end{center}
\caption{Percolation phase diagrams of self-exciting temporal processes.
We plot the percolation strength $P_\infty$ as a function of the resolution parameter $\Delta$ for
various configurations of the rate of Eq.~(\ref{eq:rate}),
with exponential kernel function and various system sizes $K$. 
Average values are obtained by considering $R = 10^3$ realizations of the process. (a) We set
$n=0$ and $\mu=1$.
The inset shows the same data as of the main with abscissa rescaled as $\Delta / \Delta^*$. (b) We
set $n=1$ and $\mu = 10^{-4}$. The insets 
display the same data as of the main, but with rescaled abscissa, 
$\Delta / \Delta_1^*$ in the lower
inset, and $\Delta / \Delta_2^*$in the upper inset. 
(c) We set $n=1$ and 
$\mu=10^2$. The inset shows the same data as of the main, but with abscissa rescaled as  
$\Delta / \Delta^*$.}
\label{fig:1}
\end{figure*}

We address the above questions in a controlled setting where we 
generate time series via an Hawkes process~\cite{hawkes1971spectra, hawkes1974cluster} with conditional rate
\begin{equation}
    \lambda(t|t_1, \ldots, t_k) = \mu + n \, \sum_{i=1}^k \, \phi(t-t_i) \; .
    \label{eq:rate}
\end{equation}
The rate  depends on the $k$ earlier 
events happened at 
times $t_1 \leq t_2 \leq \ldots \leq t_k \leq t$. 
The first term in Eq.~\eqref{eq:rate} produces spontaneous events at rate $\mu \geq 0$. 
The second term consists of the sum of individual contributions from
each earlier event, with the $i$-th event happened at time $t_i \leq
t$ increasing the rate by
$\phi(t-t_i)$.  
$\phi(x)$ is the excitation or kernel function of the self-exciting process, and it is assumed to be non-negative and monotonically non-increasing.
Typical choices for the kernel are exponential or power-law decaying functions. We 
will  consider both cases. In Eq.~(\ref{eq:rate}), we assume 
$\int_{0}^\infty \,  \phi(x) \, dx = 1$, so that the memory term is weighted by 
the single parameter $n \geq 0$. Unless otherwise stated, we always set $n=1$, 
corresponding to the critical dynamical regime of the temporal point process 
described by Eq.~(\ref{eq:rate})~\cite{helmstetter2002subcritical}.

The percolation framework allows us to characterize the generic Hawkes process 
of Eq.~(\ref{eq:rate}) using finite-size scaling analysis~\cite{stauffer2018introduction} (SM, sec.~C).
The total number $K$ of events in the time series is the system
size. For a given value of $K$,
we generate multiple time series and compute 
the percolation strength $P_\infty$, i.e.,  
the fraction of events belonging to the largest cluster,
and the associated susceptibility (SM, sec.~C).
By studying the behavior of these macroscopic observables as $K$ grows, 
we estimate the values of the
thresholds and the critical exponents.

%phenomenology
%n = 0, homogeneous Poisson process

Let us start with the case
$n = 0$ (Figure~\ref{fig:1}a), describing a homogeneous Poisson process with rate $\mu$. The generic 
inter-event time $x_i = t_i - t_{i-1}$ is a random variate distributed as $P(x_i) = \mu \, e^{-\mu \, x_i}$. Two consecutive events are part of the
same cluster with probability $P(x_i \leq \Delta) = 1 - e^{- \mu \Delta}$, which is independent
of the index $i$ and represents an effective bond occupation probability in a
homogeneous 1D percolation model~\cite{stauffer1978critical, stauffer2018introduction}. For finite $K$
values, $P_\infty$ sharply grows from $0$ to $1$ around the 
pseudo-critical point $\Delta^*(K) = \log (K) / \mu$ (SM, sec.~D). 
Finite-size scaling analysis indicates that the transition is
discontinuous, as expected for 1D ordinary
percolation~\cite{stauffer2018introduction}. We note that the
distributions of cluster size $P(S)$ and duration  $P(T)$ are exactly
described by the 1D percolation theory~\cite{stauffer2018introduction} (SM, sec.~D). 
They are
the product of a power-law function 
%[$P(S) \sim S^{-\tau}$ and $P(T) \sim T^{-\alpha}$, with $\tau = \alpha = 2$]
and a fast-decaying scaling function accounting for the system finite size~\cite{stauffer1978critical}. 
In this specific case, the scaling functions contain a multiplicative 
%quadratic 
term that exactly cancels the power-law term of the distribution. Therefore, the distributions have exponential behavior at $\Delta=\Delta^*$.
A clear signature of criticality is manifest in the relation between size and duration, $\langle S \rangle \sim T$, in agreement with the relation $\langle S \rangle \sim T^{(\alpha-1)/(\tau -1)}$ (SM, sec.~D).

%phi(x) = e^{-x}, n = 1
%A. mu << 1

We now consider
the Hawkes process of Eq.~\eqref{eq:rate} with
exponential kernel $\phi(x) = e^{-x}$~\cite{hawkes1974cluster, dassios2013exact}. 
Results of our finite-size scaling analysis are reported in 
Figures~\ref{fig:1}b and ~\ref{fig:1}c, for $\mu \ll 1$ and $\mu \gg 1$, respectively.

\begin{figure}[!htb]
\begin{center}
\includegraphics[width=0.45\textwidth]{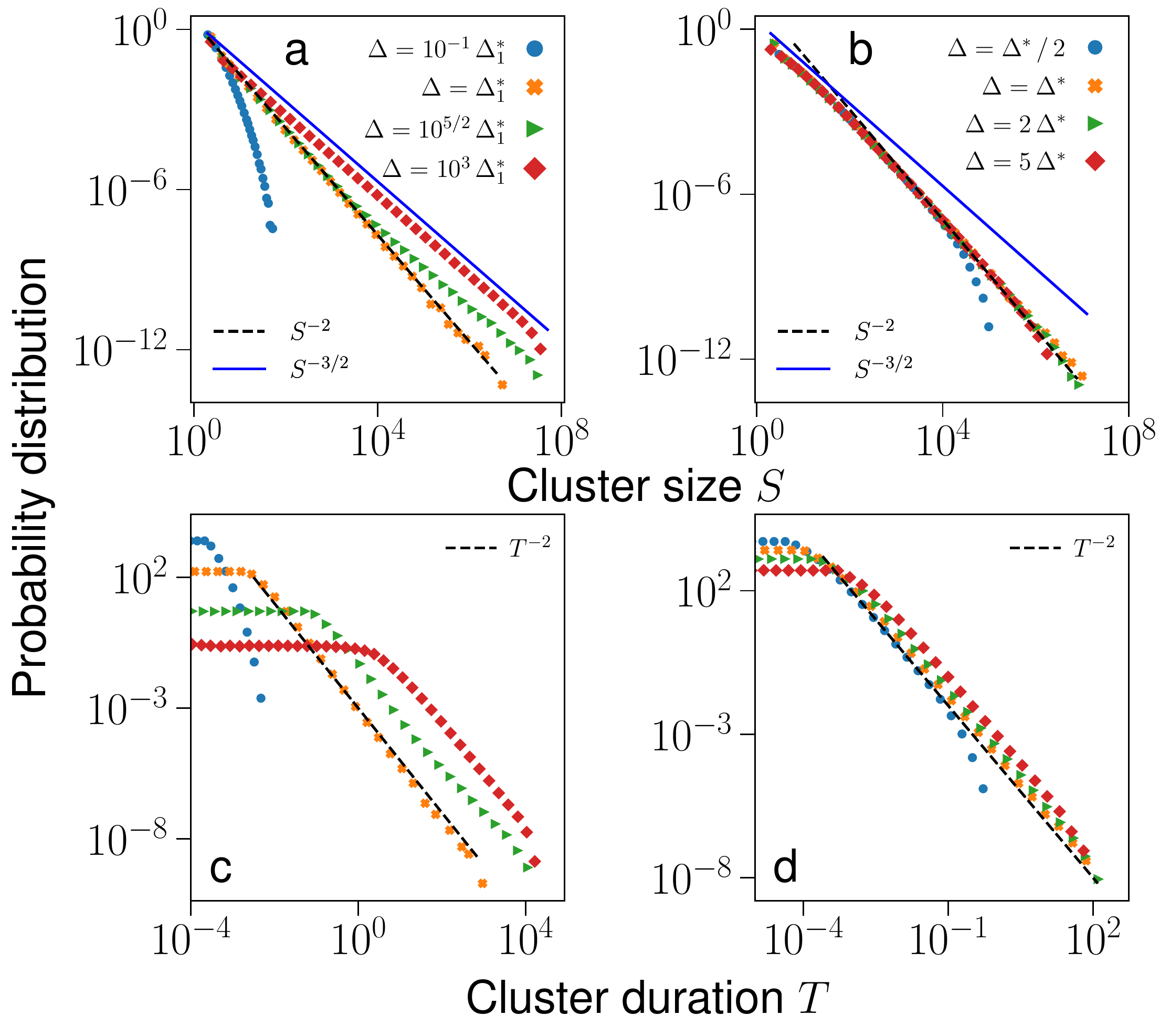}    
\end{center}
\caption{Critical properties of self-exciting temporal processes.
We consider processes 
generated from the rate of Eq.~(\ref{eq:rate}) with exponential kernel and $n=1$. 
System size $K = 10^8$. Histograms obtained by considering $C=10^7$ clusters per configuration.
(a) Cluster size distribution
for $\mu=10^{-4}$.  
(b) Cluster size distribution
for $\mu=10^2$. 
(c) Cluster duration distribution for the same data as in panel a. (d) 
Cluster duration distribution for the same data as in panel b.}
\label{fig:2}
\end{figure}

\changejan{
For $\mu \ll 1$, the phenomenology 
is rich, with two distinct transitions at $\Delta^{*}_1 < \Delta^{*}_2$, respectively.
Around the critical point $\Delta_1^*$, the system is characterized by a
behavior compatible with the universality class of 1D percolation, i.e, the same as
of the homogeneous Poisson process. Both $P(S)$ and  $P(T)$ display power-law decays at $\Delta_1^*$, with 
exponent values $\tau = \alpha =2$ (Figures~\ref{fig:2}a and~\ref{fig:2}c).
Average size and duration of clusters are linearly correlated (SM,
sec.~E). The pseudo-critical threshold equals
$\Delta^*_1 (K) \simeq \log(K)/\av{\lambda} = \log(K) / \left(\mu +\sqrt{2K\mu}\right)$, thus leading
to a vanishing critical point in the thermodynamic limit (SM, sec.~E). 
$\av{\lambda}$ is the expectation value, over an infinite number of realizations of the process, of the rate after $K$ events have happened; the estimate of the critical point $\Delta^*_1 (K)$ is thus obtained using the same exact equation as for a homogeneous Poisson process with effective rate $\av{\lambda}$.
The other transition at $\Delta^{*}_2 (K)=\log(K)/\mu$, which tends to infinite as $K$ grows,  
corresponds to the merger of the whole time series into
one cluster; its features are compatible with those of the universality class of the mean-field branching 
process, i.e., $\tau = 3/2$ and $\alpha =2$. The region of the phase diagram $[\Delta^{*}_1 (K), \Delta^{*}_2
(K)]$, which is expanding as $K$ increases, is characterized by critical behavior. 
While the percolation strength plateaus at 
$P_\infty \simeq 1 - 1/e \simeq 0.63$, the susceptibility is larger than zero.  
Furthermore, the distribution $P(S)$ displays a neat crossover between the regime $\tau = 2$ for
small $S$ and the regime $\tau = 3/2$ at large $S$ (Figure~\ref{fig:2}a).
}

%B. mu >> 1
For $\mu \gg 1$, the phase diagram displays a single transition
(Figure~\ref{fig:1}c), with 
features identical to those described for the case $\mu \ll 1$
around $\Delta_1^*$:
no crossover is present, and the critical exponents of the
distributions $P(S)$ and $P(T)$ are $\tau = \alpha =2$ (Figures~\ref{fig:2}b and ~\ref{fig:2}d).
\changejan{The same exact behavior can be obtained by simply considering a non-homogeneous Poisson process with rate linearly growing in time, i.e., $\lambda(t) \sim t$ (SM, Sec. K).}

%%%%%%%%%
The two different behaviors observed for $\mu \ll 1$ and $\mu \gg 1$ are
interpreted in an unified framework as follows. 
For $\mu \ll 1$, the process is characterized by a sequence of self-exciting bursts
due to the memory term of the rate of Eq.~(\ref{eq:rate}). Memory 
decays exponentially fast, with a typical time scale equal to 1. Each 
burst is started by a spontaneous event. 
Since spontaneous events are characterized by 
the time scale $1/\mu \gg 1$,
consecutive bursts are well separated one from the other. Increasing $\Delta$, the system exhibits first a transition "within bursts" at 
$\Delta = \Delta_1^*$, corresponding to the merger of events within the same burst, and then a transition "across bursts" at $\Delta = \Delta_2^*$, corresponding to the merger of consecutive 
bursts of activity. 
For $\mu \gg 1$, all events belong to a unique burst of self-excitation. 
The time scale of spontaneous activity is 
equal or smaller than the one due to 
self-excitation.
Thus, although the memory decays exponentially fast,
a new spontaneous event re-excites the process quickly enough
to allow the burst to proceed its activity uninterrupted.
The burst is truncated in the simulations due to the fixed size $K$ of the 
time series. As $\Delta$ increases, all events of the single burst are merged into a single 
cluster. The transition is therefore of the same type as the one observed within bursts 
at $\Delta = \Delta_1^*$ in the case $\mu \ll 1$.

%%%%%%%%%

\begin{figure}[!htb]
\begin{center}
\includegraphics[width=0.45\textwidth]{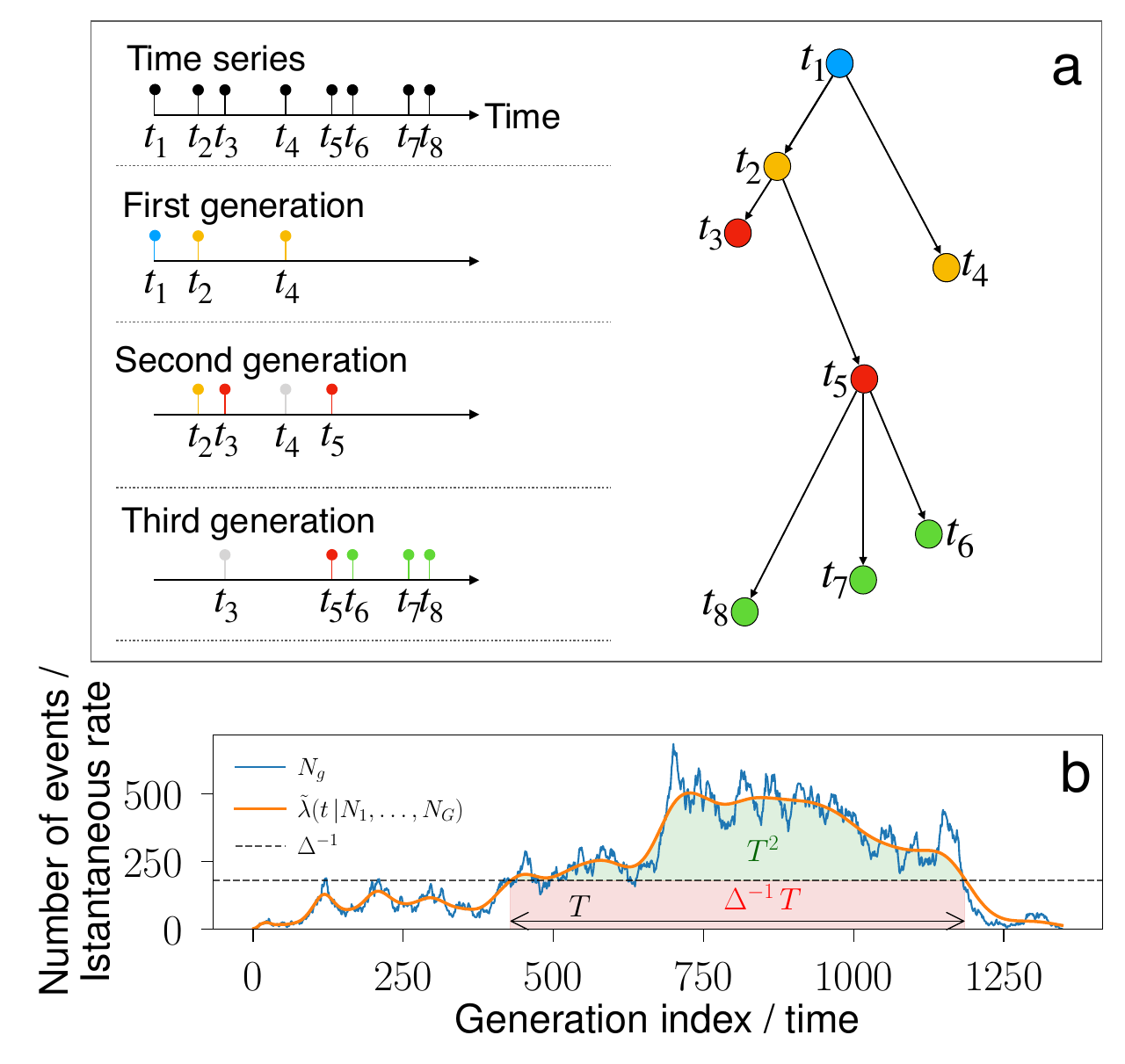}    
\end{center}
\caption{Latent tree structure of self-exciting temporal processes. (a) Each 
event in the time series on the left is associated to a parent node. On the right, the branching 
tree corresponding to the time series. Each node is assigned to a generation, and each
bond has associated an inter-event time. If $N_g$ is the number of nodes in the $g$-th generation, the depicted tree has $\{N_1=1, N_2=2, N_3=2, N_4=3, \ldots\}$. (b) The mapping of panel a allows us to  
associate to the tree $\{N_1, \dots, N_G\}$ (blue curve) an inhomogeneous Poisson process with instantaneous rate $\tilde{\lambda}(t|N_1, \ldots, N_G)$ (orange). Such a process generates time series statistically equivalent to those generated by an Hawkes process with latent tree structure $\{N_1, \ldots, N_G\}$. The inverse resolution parameter $\Delta^{-1}$ (dashed black line) is an effective threshold for the Poisson process $\tilde{\lambda}(t|N_1, \ldots, N_G)$. As a result, size and duration of clusters are related by Eq.~(\ref{eq:scaling_argument}). The shaded areas denote the two terms appearing on the rhs of Eq.~(\ref{eq:scaling_argument}).
}
\label{fig:3}
\end{figure}

%Theory of BP
We can separately study the
transitions within and across bursts. 
To this end, 
we simplify the actual process of Eq.~(\ref{eq:rate}) by setting $\mu = 0$ and assuming
that the first event of the burst already happened. We then invoke the known mapping of the 
self-exciting process of Eq.~(\ref{eq:rate}) to a standard Galton-Walton branching process 
(BP)~\cite{hawkes1974cluster}. 
According to it, the first event of the time series represents the root
of a branching tree (Figure~\ref{fig:3}). Each event generates a number
of follow-up events (offsprings) 
obeying a Poisson 
distribution with expected value equal to $n$,  
the parameter appearing in Eq.~(\ref{eq:rate}).
Time is assigned as follows. The first event happens at an arbitrary time $t_1$,
say for simplicity $t_1=0$. Then each of the following events
has associated a time equal to the time of its parent plus a random variate $x$ extracted
from the kernel function $\phi(x)$ of Eq.~(\ref{eq:rate}). 
The mapping to the BP offers an alternative \changejan{(on average statistically equivalent)} way of generating
time series for the self-exciting process of Eq.~(\ref{eq:rate}). 
We first generate a BP tree, and then associate a time to
each event of the tree according to the rule described above. 
The time $t$ associated to a generic event of the $g$-th generation is distributed according
to a function $P(t| g)$. For the exponential kernel function, 
$P(t| g)$ is the sum of $g$ exponentially distributed variables, i.e.,
the Erlang distribution with rate equal to 1, $P(t| g) = t^{g-1} \, e^{-t} / (g-1)!$.

The mapping of the self-exciting process to a BP allows us to fully understand the numerical 
findings of Figures~\ref{fig:1} and~\ref{fig:2}.
For $n=1$ the BP is critical. The distribution of the tree size is $P(Z) \sim Z^{-3/2}$
and the distribution of the tree depth is $P(D) \sim D^{-2}$. Individual bursts of
activity, as seen for sufficiently high $\Delta$ values and $\mu \ll 1$, obey 
this statistics. Specifically, the size of each burst $S$ is exactly the size 
$Z$ of the tree. The average duration of the bursts $\langle T \rangle \sim D$, as 
expected for the sum of iid exponentially distributed random variates.
For  $\Delta \in [\Delta_1^*, \Delta_2^*]$, $P_\infty$ of Figure~\ref{fig:1}b 
follows the same statistics as the maximum value of a sample of variables extracted 
from the distribution  $P(Z) \sim Z^{-3/2}$ divided by their sum, and the average value of the ratio 
plateaus at $1 - 1/e$ for sufficiently large sample sizes, fully
explaining the results of Figure~\ref{fig:1} (SM, sec.~H).

The behavior at $\Delta=\Delta_1^*$
and the crossover towards the standard BP
regime for larger $\Delta$ are due to a threshold phenomenon.
This directly follows from the abrupt nature of the percolation
transition of the Poisson process (Figure~\ref{fig:1}a). 
Given the latent branching tree $\{N_1, N_2, \ldots, N_g, \ldots,
N_G\}$, where $N_g$ indicates the number of events of the $g$-th
generation of the tree, the time series of the Hawkes process is
statistically equivalent to the one of the inhomogeneous Poisson
process with instantaneous rate $\tilde{\lambda} (t|N_1, \ldots, N_G) = \sum_g
\, P(t| g) \, N_g$. Hence, for a given $\Delta$,
as long as $\tilde{\lambda}(t|N_1, \ldots, N_G) > 1/ \Delta$,
all events are part of the same cluster of activity; when
instead $\tilde{\lambda}(t|N_1, \ldots, N_G) < 1/ \Delta$, then events around
time $t$ belong to separate clusters of activity.
As a consequence, the total number of events $S_T$ that form a cluster of
activity of duration $T$ is the integral of the curve
$\tilde{\lambda}(t|N_1, \ldots, N_G)$ in the time interval when the rate is
above $\Delta^{-1}$ (Figure~\ref{fig:3}b).
\changejan{We repeat a similar calculation as in Ref.~\cite{villegas2019time}.}
The integral can be split in two contributions, %~\cite{villegas2019time},
one corresponding to the area of the order of $T^2$ appearing above
the threshold line, as expected for a critical BP~\cite{di2017simple,
villegas2019time}, and the other corresponding to the
area $\Delta^{-1} \, T$ appearing below threshold,
\begin{equation}
S_T \sim T^2 + \Delta^{-1} \, T \; .  \label{eq:scaling_argument}
\end{equation}
While the distribution of cluster durations is always the same
[i.e., $P(T) \sim T^{-2}$ of the underlying BP], 
if $\Delta^{-1}>T$ then $S_T \sim T$ implying the
within-burst statistics $P(S) \sim S^{-2}$. Instead, if $\Delta^{-1}<T$
then $S_T \sim T^2$ and the
conservation of probability leads to
the BP statistics $P(S) \sim S^{-3/2}$.
When the two terms on the rhs of
Eq.~(\ref{eq:scaling_argument}) have comparable magnitude,
a crossover between the two scalings occurs. 
The crossover point varies with the temporal resolution as $S_c
\propto \Delta^{-2}$ (SM, sec.~G). A full
understanding of $P(S)$ is achieved by noting that power-law scaling
requires a minimum sample size to be observed, sufficient for the largest cluster to have duration
comparable to $1/\Delta_1^*$. If the sample is not large enough the
distribution will appear as exponential (SM, sec.~G).

% phi(x) = (gamma-1) (1 + x)^{-gamma}
We finally consider the power-law kernel function 
$\phi(x) = (\gamma-1) (1 + x)^{-\gamma}$. The branching structure underlying 
the process is not affected by the kernel so the 
results above should continue to hold~\cite{helmstetter2002subcritical}. 
For $\gamma > 2$, $\phi(x)$ has finite mean value and, as a consequence, results 
are identical to those obtained for the exponential kernel (SM, Sec.~G).
Specifically, $P_{\infty}$ shows a discontinuous transition
when $\mu \gg 1$, while two sharp transitions are 
observed for $\mu \ll 1$. The distribution of cluster sizes exhibits a crossover from $\tau=2$ at 
$\Delta^*_1$ to $\tau=3/2$ for $\Delta \gg \Delta_1^*$ when $\mu \ll 1$, and the 
exponent $\tau=2$ with no crossover when $\mu \gg 1$. 
If $\gamma \leq 2$, $\phi(x)$ has diverging mean value, the  typical inter-event time is 
large preventing the present framework to be applicable.

In summary, we investigated how self-excitation mechanisms are reflected in the bursty dynamics, exploring their relationship with avalanche distributions, which offer an effective probe into the presence of autocorrelation in time series~\cite{karsai2018bursty}.
%We focused on the Hawkes process, a general mechanism to produce self-excitation and autocorrelation, simple enough to allow semi-rigorous treatment. We found that the presence of self-excitation mechanisms produces fat-tailed distributions in the avalanche size and duration distributions.
\change{We focused on the Hawkes process, a general mechanism to produce self-excitation, autocorrelation, and fat-tailed distributions in the avalanche size and duration.}
Critical behavior in the distributions is observed at specific values of the resolution parameter $\Delta$, and is characterized by exponents independent of the form of the self-excitation mechanism. The  universal critical behavior is governed by both the branching structure underlying the Hawkes process and the features of 1D percolation.  Nontrivial details of the size distribution depend on the relative force of the spontaneous and self-excitation mechanisms. The two classes of behavior coexist for a wide range of $\Delta$ values, thus making the observation of a mixture of two classes the most likely outcome of an analysis where the resolution parameter is not fine-tuned. 
\change{All findings extend to the slightly subcritical 
configuration of the Hawkes process (\SM, Sec.~I), thus showing that our method is scientifically sound also for the analysis of avalanches in some natural systems possibly operating close to, but not exactly in, a critical regime~\cite{priesemann2014spike}.} 
Our work offers an interpretative framework for the relationship between avalanche properties and the mechanisms producing autocorrelation in bursty dynamics.
More work in this area is nevertheless needed. The Hawkes process is
unable to reproduce the variety of critical behaviors reported for
real data sets in Ref.~\cite{karsai2018bursty}, and other
self-excitation mechanisms need to be considered.

\acknowledgements{
D.N. and F.R. acknowledge partial support from the National Science Foundation (Grant No. CMMI-1552487).
  }

\bibliography{biblio}

\begin{thebibliography}{36}%
\makeatletter
\providecommand \@ifxundefined [1]{%
 \@ifx{#1\undefined}
}%
\providecommand \@ifnum [1]{%
 \ifnum #1\expandafter \@firstoftwo
 \else \expandafter \@secondoftwo
 \fi
}%
\providecommand \@ifx [1]{%
 \ifx #1\expandafter \@firstoftwo
 \else \expandafter \@secondoftwo
 \fi
}%
\providecommand \natexlab [1]{#1}%
\providecommand \enquote  [1]{``#1''}%
\providecommand \bibnamefont  [1]{#1}%
\providecommand \bibfnamefont [1]{#1}%
\providecommand \citenamefont [1]{#1}%
\providecommand \href@noop [0]{\@secondoftwo}%
\providecommand \href [0]{\begingroup \@sanitize@url \@href}%
\providecommand \@href[1]{\@@startlink{#1}\@@href}%
\providecommand \@@href[1]{\endgroup#1\@@endlink}%
\providecommand \@sanitize@url [0]{\catcode `\\12\catcode `\$12\catcode
  `\&12\catcode `\#12\catcode `\^12\catcode `\_12\catcode `\%12\relax}%
\providecommand \@@startlink[1]{}%
\providecommand \@@endlink[0]{}%
\providecommand \url  [0]{\begingroup\@sanitize@url \@url }%
\providecommand \@url [1]{\endgroup\@href {#1}{\urlprefix }}%
\providecommand \urlprefix  [0]{URL }%
\providecommand \Eprint [0]{\href }%
\providecommand \doibase [0]{http://dx.doi.org/}%
\providecommand \selectlanguage [0]{\@gobble}%
\providecommand \bibinfo  [0]{\@secondoftwo}%
\providecommand \bibfield  [0]{\@secondoftwo}%
\providecommand \translation [1]{[#1]}%
\providecommand \BibitemOpen [0]{}%
\providecommand \bibitemStop [0]{}%
\providecommand \bibitemNoStop [0]{.\EOS\space}%
\providecommand \EOS [0]{\spacefactor3000\relax}%
\providecommand \BibitemShut  [1]{\csname bibitem#1\endcsname}%
\let\auto@bib@innerbib\@empty
%</preamble>
\bibitem [{\citenamefont {Karsai}\ \emph {et~al.}(2018)\citenamefont {Karsai},
  \citenamefont {Jo},\ and\ \citenamefont {Kaski}}]{karsai2018bursty}%
  \BibitemOpen
  \bibfield  {author} {\bibinfo {author} {\bibfnamefont {M.}~\bibnamefont
  {Karsai}}, \bibinfo {author} {\bibfnamefont {H.-H.}\ \bibnamefont {Jo}}, \
  and\ \bibinfo {author} {\bibfnamefont {K.}~\bibnamefont {Kaski}},\
  }\href@noop {} {\emph {\bibinfo {title} {Bursty human dynamics}}}\ (\bibinfo
  {publisher} {Springer},\ \bibinfo {year} {2018})\BibitemShut {NoStop}%
\bibitem [{\citenamefont {Dalla~Porta}\ and\ \citenamefont
  {Copelli}(2019)}]{dalla2019modeling}%
  \BibitemOpen
  \bibfield  {author} {\bibinfo {author} {\bibfnamefont {L.}~\bibnamefont
  {Dalla~Porta}}\ and\ \bibinfo {author} {\bibfnamefont {M.}~\bibnamefont
  {Copelli}},\ }\href@noop {} {\bibfield  {journal} {\bibinfo  {journal} {PLoS
  computational biology}\ }\textbf {\bibinfo {volume} {15}},\ \bibinfo {pages}
  {e1006924} (\bibinfo {year} {2019})}\BibitemShut {NoStop}%
\bibitem [{\citenamefont {Beggs}\ and\ \citenamefont
  {Plenz}(2003)}]{beggs2003neuronal}%
  \BibitemOpen
  \bibfield  {author} {\bibinfo {author} {\bibfnamefont {J.~M.}\ \bibnamefont
  {Beggs}}\ and\ \bibinfo {author} {\bibfnamefont {D.}~\bibnamefont {Plenz}},\
  }\href@noop {} {\bibfield  {journal} {\bibinfo  {journal} {Journal of
  neuroscience}\ }\textbf {\bibinfo {volume} {23}},\ \bibinfo {pages} {11167}
  (\bibinfo {year} {2003})}\BibitemShut {NoStop}%
\bibitem [{\citenamefont {Bak}\ \emph {et~al.}(2002)\citenamefont {Bak},
  \citenamefont {Christensen}, \citenamefont {Danon},\ and\ \citenamefont
  {Scanlon}}]{bak2002unified}%
  \BibitemOpen
  \bibfield  {author} {\bibinfo {author} {\bibfnamefont {P.}~\bibnamefont
  {Bak}}, \bibinfo {author} {\bibfnamefont {K.}~\bibnamefont {Christensen}},
  \bibinfo {author} {\bibfnamefont {L.}~\bibnamefont {Danon}}, \ and\ \bibinfo
  {author} {\bibfnamefont {T.}~\bibnamefont {Scanlon}},\ }\href@noop {}
  {\bibfield  {journal} {\bibinfo  {journal} {Physical Review Letters}\
  }\textbf {\bibinfo {volume} {88}},\ \bibinfo {pages} {178501} (\bibinfo
  {year} {2002})}\BibitemShut {NoStop}%
\bibitem [{\citenamefont {Wang}\ and\ \citenamefont
  {Dai}(2013)}]{wang2013self}%
  \BibitemOpen
  \bibfield  {author} {\bibinfo {author} {\bibfnamefont {F.}~\bibnamefont
  {Wang}}\ and\ \bibinfo {author} {\bibfnamefont {Z.}~\bibnamefont {Dai}},\
  }\href@noop {} {\bibfield  {journal} {\bibinfo  {journal} {Nature Physics}\
  }\textbf {\bibinfo {volume} {9}},\ \bibinfo {pages} {465} (\bibinfo {year}
  {2013})}\BibitemShut {NoStop}%
\bibitem [{\citenamefont {Barabasi}(2005)}]{barabasi2005origin}%
  \BibitemOpen
  \bibfield  {author} {\bibinfo {author} {\bibfnamefont {A.-L.}\ \bibnamefont
  {Barabasi}},\ }\href@noop {} {\bibfield  {journal} {\bibinfo  {journal}
  {Nature}\ }\textbf {\bibinfo {volume} {435}},\ \bibinfo {pages} {207}
  (\bibinfo {year} {2005})}\BibitemShut {NoStop}%
\bibitem [{\citenamefont {Karsai}\ \emph {et~al.}(2012)\citenamefont {Karsai},
  \citenamefont {Kaski}, \citenamefont {Barab{\'a}si},\ and\ \citenamefont
  {Kert{\'e}sz}}]{karsai2012universal}%
  \BibitemOpen
  \bibfield  {author} {\bibinfo {author} {\bibfnamefont {M.}~\bibnamefont
  {Karsai}}, \bibinfo {author} {\bibfnamefont {K.}~\bibnamefont {Kaski}},
  \bibinfo {author} {\bibfnamefont {A.-L.}\ \bibnamefont {Barab{\'a}si}}, \
  and\ \bibinfo {author} {\bibfnamefont {J.}~\bibnamefont {Kert{\'e}sz}},\
  }\href@noop {} {\bibfield  {journal} {\bibinfo  {journal} {Scientific
  reports}\ }\textbf {\bibinfo {volume} {2}},\ \bibinfo {pages} {1} (\bibinfo
  {year} {2012})}\BibitemShut {NoStop}%
\bibitem [{\citenamefont {Gleeson}\ \emph {et~al.}(2014)\citenamefont
  {Gleeson}, \citenamefont {Ward}, \citenamefont {O’sullivan},\ and\
  \citenamefont {Lee}}]{gleeson2014competition}%
  \BibitemOpen
  \bibfield  {author} {\bibinfo {author} {\bibfnamefont {J.~P.}\ \bibnamefont
  {Gleeson}}, \bibinfo {author} {\bibfnamefont {J.~A.}\ \bibnamefont {Ward}},
  \bibinfo {author} {\bibfnamefont {K.~P.}\ \bibnamefont {O’sullivan}}, \
  and\ \bibinfo {author} {\bibfnamefont {W.~T.}\ \bibnamefont {Lee}},\
  }\href@noop {} {\bibfield  {journal} {\bibinfo  {journal} {Physical review
  letters}\ }\textbf {\bibinfo {volume} {112}},\ \bibinfo {pages} {048701}
  (\bibinfo {year} {2014})}\BibitemShut {NoStop}%
\bibitem [{\citenamefont {Fontenele}\ \emph {et~al.}(2019)\citenamefont
  {Fontenele}, \citenamefont {de~Vasconcelos}, \citenamefont {Feliciano},
  \citenamefont {Aguiar}, \citenamefont {Soares-Cunha}, \citenamefont
  {Coimbra}, \citenamefont {Dalla~Porta}, \citenamefont {Ribeiro},
  \citenamefont {Rodrigues}, \citenamefont {Sousa} \emph
  {et~al.}}]{fontenele2019criticality}%
  \BibitemOpen
  \bibfield  {author} {\bibinfo {author} {\bibfnamefont {A.~J.}\ \bibnamefont
  {Fontenele}}, \bibinfo {author} {\bibfnamefont {N.~A.}\ \bibnamefont
  {de~Vasconcelos}}, \bibinfo {author} {\bibfnamefont {T.}~\bibnamefont
  {Feliciano}}, \bibinfo {author} {\bibfnamefont {L.~A.}\ \bibnamefont
  {Aguiar}}, \bibinfo {author} {\bibfnamefont {C.}~\bibnamefont
  {Soares-Cunha}}, \bibinfo {author} {\bibfnamefont {B.}~\bibnamefont
  {Coimbra}}, \bibinfo {author} {\bibfnamefont {L.}~\bibnamefont
  {Dalla~Porta}}, \bibinfo {author} {\bibfnamefont {S.}~\bibnamefont
  {Ribeiro}}, \bibinfo {author} {\bibfnamefont {A.~J.}\ \bibnamefont
  {Rodrigues}}, \bibinfo {author} {\bibfnamefont {N.}~\bibnamefont {Sousa}},
  \emph {et~al.},\ }\href@noop {} {\bibfield  {journal} {\bibinfo  {journal}
  {Physical review letters}\ }\textbf {\bibinfo {volume} {122}},\ \bibinfo
  {pages} {208101} (\bibinfo {year} {2019})}\BibitemShut {NoStop}%
\bibitem [{\citenamefont {Weng}\ \emph {et~al.}(2012)\citenamefont {Weng},
  \citenamefont {Flammini}, \citenamefont {Vespignani},\ and\ \citenamefont
  {Menczer}}]{weng2012competition}%
  \BibitemOpen
  \bibfield  {author} {\bibinfo {author} {\bibfnamefont {L.}~\bibnamefont
  {Weng}}, \bibinfo {author} {\bibfnamefont {A.}~\bibnamefont {Flammini}},
  \bibinfo {author} {\bibfnamefont {A.}~\bibnamefont {Vespignani}}, \ and\
  \bibinfo {author} {\bibfnamefont {F.}~\bibnamefont {Menczer}},\ }\href@noop
  {} {\bibfield  {journal} {\bibinfo  {journal} {Scientific reports}\ }\textbf
  {\bibinfo {volume} {2}},\ \bibinfo {pages} {335} (\bibinfo {year}
  {2012})}\BibitemShut {NoStop}%
\bibitem [{\citenamefont {Jo}\ \emph {et~al.}(2015)\citenamefont {Jo},
  \citenamefont {Perotti}, \citenamefont {Kaski},\ and\ \citenamefont
  {Kert{\'e}sz}}]{jo2015correlated}%
  \BibitemOpen
  \bibfield  {author} {\bibinfo {author} {\bibfnamefont {H.-H.}\ \bibnamefont
  {Jo}}, \bibinfo {author} {\bibfnamefont {J.~I.}\ \bibnamefont {Perotti}},
  \bibinfo {author} {\bibfnamefont {K.}~\bibnamefont {Kaski}}, \ and\ \bibinfo
  {author} {\bibfnamefont {J.}~\bibnamefont {Kert{\'e}sz}},\ }\href@noop {}
  {\bibfield  {journal} {\bibinfo  {journal} {Physical Review E}\ }\textbf
  {\bibinfo {volume} {92}},\ \bibinfo {pages} {022814} (\bibinfo {year}
  {2015})}\BibitemShut {NoStop}%
\bibitem [{\citenamefont {Kumar}\ \emph {et~al.}(2020)\citenamefont {Kumar},
  \citenamefont {Korkolis}, \citenamefont {Benzi}, \citenamefont {Denisov},
  \citenamefont {Niemeijer}, \citenamefont {Schall}, \citenamefont {Toschi},\
  and\ \citenamefont {Trampert}}]{kumar2020interevent}%
  \BibitemOpen
  \bibfield  {author} {\bibinfo {author} {\bibfnamefont {P.}~\bibnamefont
  {Kumar}}, \bibinfo {author} {\bibfnamefont {E.}~\bibnamefont {Korkolis}},
  \bibinfo {author} {\bibfnamefont {R.}~\bibnamefont {Benzi}}, \bibinfo
  {author} {\bibfnamefont {D.}~\bibnamefont {Denisov}}, \bibinfo {author}
  {\bibfnamefont {A.}~\bibnamefont {Niemeijer}}, \bibinfo {author}
  {\bibfnamefont {P.}~\bibnamefont {Schall}}, \bibinfo {author} {\bibfnamefont
  {F.}~\bibnamefont {Toschi}}, \ and\ \bibinfo {author} {\bibfnamefont
  {J.}~\bibnamefont {Trampert}},\ }\href@noop {} {\bibfield  {journal}
  {\bibinfo  {journal} {Scientific reports}\ }\textbf {\bibinfo {volume}
  {10}},\ \bibinfo {pages} {1} (\bibinfo {year} {2020})}\BibitemShut {NoStop}%
\bibitem [{\citenamefont {Pasquale}\ \emph {et~al.}(2008)\citenamefont
  {Pasquale}, \citenamefont {Massobrio}, \citenamefont {Bologna}, \citenamefont
  {Chiappalone},\ and\ \citenamefont {Martinoia}}]{pasquale2008self}%
  \BibitemOpen
  \bibfield  {author} {\bibinfo {author} {\bibfnamefont {V.}~\bibnamefont
  {Pasquale}}, \bibinfo {author} {\bibfnamefont {P.}~\bibnamefont {Massobrio}},
  \bibinfo {author} {\bibfnamefont {L.}~\bibnamefont {Bologna}}, \bibinfo
  {author} {\bibfnamefont {M.}~\bibnamefont {Chiappalone}}, \ and\ \bibinfo
  {author} {\bibfnamefont {S.}~\bibnamefont {Martinoia}},\ }\href@noop {}
  {\bibfield  {journal} {\bibinfo  {journal} {Neuroscience}\ }\textbf {\bibinfo
  {volume} {153}},\ \bibinfo {pages} {1354} (\bibinfo {year}
  {2008})}\BibitemShut {NoStop}%
\bibitem [{\citenamefont {Neto}\ \emph {et~al.}(2019)\citenamefont {Neto},
  \citenamefont {Spitzner},\ and\ \citenamefont
  {Priesemann}}]{neto2019unified}%
  \BibitemOpen
  \bibfield  {author} {\bibinfo {author} {\bibfnamefont {J.~P.}\ \bibnamefont
  {Neto}}, \bibinfo {author} {\bibfnamefont {F.~P.}\ \bibnamefont {Spitzner}},
  \ and\ \bibinfo {author} {\bibfnamefont {V.}~\bibnamefont {Priesemann}},\
  }\href@noop {} {\bibfield  {journal} {\bibinfo  {journal} {arXiv preprint
  arXiv:1910.09984}\ } (\bibinfo {year} {2019})}\BibitemShut {NoStop}%
\bibitem [{\citenamefont {Miller}\ \emph {et~al.}(2019)\citenamefont {Miller},
  \citenamefont {Yu},\ and\ \citenamefont {Plenz}}]{miller2019scale}%
  \BibitemOpen
  \bibfield  {author} {\bibinfo {author} {\bibfnamefont {S.~R.}\ \bibnamefont
  {Miller}}, \bibinfo {author} {\bibfnamefont {S.}~\bibnamefont {Yu}}, \ and\
  \bibinfo {author} {\bibfnamefont {D.}~\bibnamefont {Plenz}},\ }\href@noop {}
  {\bibfield  {journal} {\bibinfo  {journal} {Scientific reports}\ }\textbf
  {\bibinfo {volume} {9}},\ \bibinfo {pages} {1} (\bibinfo {year}
  {2019})}\BibitemShut {NoStop}%
\bibitem [{\citenamefont {Chiappalone}\ \emph {et~al.}(2005)\citenamefont
  {Chiappalone}, \citenamefont {Novellino}, \citenamefont {Vajda},
  \citenamefont {Vato}, \citenamefont {Martinoia},\ and\ \citenamefont {van
  Pelt}}]{chiappalone2005burst}%
  \BibitemOpen
  \bibfield  {author} {\bibinfo {author} {\bibfnamefont {M.}~\bibnamefont
  {Chiappalone}}, \bibinfo {author} {\bibfnamefont {A.}~\bibnamefont
  {Novellino}}, \bibinfo {author} {\bibfnamefont {I.}~\bibnamefont {Vajda}},
  \bibinfo {author} {\bibfnamefont {A.}~\bibnamefont {Vato}}, \bibinfo {author}
  {\bibfnamefont {S.}~\bibnamefont {Martinoia}}, \ and\ \bibinfo {author}
  {\bibfnamefont {J.}~\bibnamefont {van Pelt}},\ }\href@noop {} {\bibfield
  {journal} {\bibinfo  {journal} {Neurocomputing}\ }\textbf {\bibinfo {volume}
  {65}},\ \bibinfo {pages} {653} (\bibinfo {year} {2005})}\BibitemShut
  {NoStop}%
\bibitem [{\citenamefont {Levina}\ and\ \citenamefont
  {Priesemann}(2017)}]{levina2017subsampling}%
  \BibitemOpen
  \bibfield  {author} {\bibinfo {author} {\bibfnamefont {A.}~\bibnamefont
  {Levina}}\ and\ \bibinfo {author} {\bibfnamefont {V.}~\bibnamefont
  {Priesemann}},\ }\href@noop {} {\bibfield  {journal} {\bibinfo  {journal}
  {Nature communications}\ }\textbf {\bibinfo {volume} {8}},\ \bibinfo {pages}
  {1} (\bibinfo {year} {2017})}\BibitemShut {NoStop}%
\bibitem [{\citenamefont {Jani{\'c}evi{\'c}}\ \emph {et~al.}(2016)\citenamefont
  {Jani{\'c}evi{\'c}}, \citenamefont {Laurson}, \citenamefont {M{\aa}l{\o}y},
  \citenamefont {Santucci},\ and\ \citenamefont
  {Alava}}]{janicevic2016interevent}%
  \BibitemOpen
  \bibfield  {author} {\bibinfo {author} {\bibfnamefont {S.}~\bibnamefont
  {Jani{\'c}evi{\'c}}}, \bibinfo {author} {\bibfnamefont {L.}~\bibnamefont
  {Laurson}}, \bibinfo {author} {\bibfnamefont {K.~J.}\ \bibnamefont
  {M{\aa}l{\o}y}}, \bibinfo {author} {\bibfnamefont {S.}~\bibnamefont
  {Santucci}}, \ and\ \bibinfo {author} {\bibfnamefont {M.~J.}\ \bibnamefont
  {Alava}},\ }\href@noop {} {\bibfield  {journal} {\bibinfo  {journal}
  {Physical review letters}\ }\textbf {\bibinfo {volume} {117}},\ \bibinfo
  {pages} {230601} (\bibinfo {year} {2016})}\BibitemShut {NoStop}%
\bibitem [{\citenamefont {Villegas}\ \emph {et~al.}(2019)\citenamefont
  {Villegas}, \citenamefont {di~Santo}, \citenamefont {Burioni},\ and\
  \citenamefont {Mu{\~n}oz}}]{villegas2019time}%
  \BibitemOpen
  \bibfield  {author} {\bibinfo {author} {\bibfnamefont {P.}~\bibnamefont
  {Villegas}}, \bibinfo {author} {\bibfnamefont {S.}~\bibnamefont {di~Santo}},
  \bibinfo {author} {\bibfnamefont {R.}~\bibnamefont {Burioni}}, \ and\
  \bibinfo {author} {\bibfnamefont {M.~A.}\ \bibnamefont {Mu{\~n}oz}},\
  }\href@noop {} {\bibfield  {journal} {\bibinfo  {journal} {Physical Review
  E}\ }\textbf {\bibinfo {volume} {100}},\ \bibinfo {pages} {012133} (\bibinfo
  {year} {2019})}\BibitemShut {NoStop}%
\bibitem [{\citenamefont {Hawkes}(1971)}]{hawkes1971spectra}%
  \BibitemOpen
  \bibfield  {author} {\bibinfo {author} {\bibfnamefont {A.~G.}\ \bibnamefont
  {Hawkes}},\ }\href@noop {} {\bibfield  {journal} {\bibinfo  {journal}
  {Biometrika}\ }\textbf {\bibinfo {volume} {58}},\ \bibinfo {pages} {83}
  (\bibinfo {year} {1971})}\BibitemShut {NoStop}%
\bibitem [{\citenamefont {Helmstetter}\ and\ \citenamefont
  {Sornette}(2002)}]{helmstetter2002subcritical}%
  \BibitemOpen
  \bibfield  {author} {\bibinfo {author} {\bibfnamefont {A.}~\bibnamefont
  {Helmstetter}}\ and\ \bibinfo {author} {\bibfnamefont {D.}~\bibnamefont
  {Sornette}},\ }\href@noop {} {\bibfield  {journal} {\bibinfo  {journal}
  {Journal of Geophysical Research: Solid Earth}\ }\textbf {\bibinfo {volume}
  {107}},\ \bibinfo {pages} {ESE} (\bibinfo {year} {2002})}\BibitemShut
  {NoStop}%
\bibitem [{\citenamefont {Ogata}(1988)}]{ogata1988statistical}%
  \BibitemOpen
  \bibfield  {author} {\bibinfo {author} {\bibfnamefont {Y.}~\bibnamefont
  {Ogata}},\ }\href@noop {} {\bibfield  {journal} {\bibinfo  {journal} {Journal
  of the American Statistical association}\ }\textbf {\bibinfo {volume} {83}},\
  \bibinfo {pages} {9} (\bibinfo {year} {1988})}\BibitemShut {NoStop}%
\bibitem [{\citenamefont {Kossio}\ \emph {et~al.}(2018)\citenamefont {Kossio},
  \citenamefont {Goedeke}, \citenamefont {van~den Akker}, \citenamefont
  {Ibarz},\ and\ \citenamefont {Memmesheimer}}]{kossio2018growing}%
  \BibitemOpen
  \bibfield  {author} {\bibinfo {author} {\bibfnamefont {F.~Y.~K.}\
  \bibnamefont {Kossio}}, \bibinfo {author} {\bibfnamefont {S.}~\bibnamefont
  {Goedeke}}, \bibinfo {author} {\bibfnamefont {B.}~\bibnamefont {van~den
  Akker}}, \bibinfo {author} {\bibfnamefont {B.}~\bibnamefont {Ibarz}}, \ and\
  \bibinfo {author} {\bibfnamefont {R.-M.}\ \bibnamefont {Memmesheimer}},\
  }\href@noop {} {\bibfield  {journal} {\bibinfo  {journal} {Physical review
  letters}\ }\textbf {\bibinfo {volume} {121}},\ \bibinfo {pages} {058301}
  (\bibinfo {year} {2018})}\BibitemShut {NoStop}%
\bibitem [{\citenamefont {Crane}\ and\ \citenamefont
  {Sornette}(2008)}]{crane2008robust}%
  \BibitemOpen
  \bibfield  {author} {\bibinfo {author} {\bibfnamefont {R.}~\bibnamefont
  {Crane}}\ and\ \bibinfo {author} {\bibfnamefont {D.}~\bibnamefont
  {Sornette}},\ }\href@noop {} {\bibfield  {journal} {\bibinfo  {journal}
  {Proceedings of the National Academy of Sciences}\ }\textbf {\bibinfo
  {volume} {105}},\ \bibinfo {pages} {15649} (\bibinfo {year}
  {2008})}\BibitemShut {NoStop}%
\bibitem [{\citenamefont {Sornette}\ \emph {et~al.}(2004)\citenamefont
  {Sornette}, \citenamefont {Desch{\^a}tres}, \citenamefont {Gilbert},\ and\
  \citenamefont {Ageon}}]{sornette2004endogenous}%
  \BibitemOpen
  \bibfield  {author} {\bibinfo {author} {\bibfnamefont {D.}~\bibnamefont
  {Sornette}}, \bibinfo {author} {\bibfnamefont {F.}~\bibnamefont
  {Desch{\^a}tres}}, \bibinfo {author} {\bibfnamefont {T.}~\bibnamefont
  {Gilbert}}, \ and\ \bibinfo {author} {\bibfnamefont {Y.}~\bibnamefont
  {Ageon}},\ }\href@noop {} {\bibfield  {journal} {\bibinfo  {journal}
  {Physical Review Letters}\ }\textbf {\bibinfo {volume} {93}},\ \bibinfo
  {pages} {228701} (\bibinfo {year} {2004})}\BibitemShut {NoStop}%
\bibitem [{\citenamefont {Stauffer}\ and\ \citenamefont
  {Aharony}(1994)}]{stauffer2018introduction}%
  \BibitemOpen
  \bibfield  {author} {\bibinfo {author} {\bibfnamefont {D.}~\bibnamefont
  {Stauffer}}\ and\ \bibinfo {author} {\bibfnamefont {A.}~\bibnamefont
  {Aharony}},\ }\href@noop {} {\emph {\bibinfo {title} {Introduction to
  percolation theory}}}\ (\bibinfo  {publisher} {Taylor and Francis},\ \bibinfo
  {year} {1994})\BibitemShut {NoStop}%
\bibitem [{\citenamefont {Hawkes}\ and\ \citenamefont
  {Oakes}(1974)}]{hawkes1974cluster}%
  \BibitemOpen
  \bibfield  {author} {\bibinfo {author} {\bibfnamefont {A.~G.}\ \bibnamefont
  {Hawkes}}\ and\ \bibinfo {author} {\bibfnamefont {D.}~\bibnamefont {Oakes}},\
  }\href@noop {} {\bibfield  {journal} {\bibinfo  {journal} {Journal of Applied
  Probability}\ }\textbf {\bibinfo {volume} {11}},\ \bibinfo {pages} {493}
  (\bibinfo {year} {1974})}\BibitemShut {NoStop}%
\bibitem [{\citenamefont {Stauffer}\ and\ \citenamefont
  {Jayaprakash}(1978)}]{stauffer1978critical}%
  \BibitemOpen
  \bibfield  {author} {\bibinfo {author} {\bibfnamefont {D.}~\bibnamefont
  {Stauffer}}\ and\ \bibinfo {author} {\bibfnamefont {C.}~\bibnamefont
  {Jayaprakash}},\ }\href@noop {} {\bibfield  {journal} {\bibinfo  {journal}
  {Physics Letters A}\ }\textbf {\bibinfo {volume} {64}},\ \bibinfo {pages}
  {433} (\bibinfo {year} {1978})}\BibitemShut {NoStop}%
\bibitem [{\citenamefont {Dassios}\ and\ \citenamefont
  {Zhao}(2013)}]{dassios2013exact}%
  \BibitemOpen
  \bibfield  {author} {\bibinfo {author} {\bibfnamefont {A.}~\bibnamefont
  {Dassios}}\ and\ \bibinfo {author} {\bibfnamefont {H.}~\bibnamefont {Zhao}},\
  }\href@noop {} {\bibfield  {journal} {\bibinfo  {journal} {Electronic
  Communications in Probability}\ }\textbf {\bibinfo {volume} {18}} (\bibinfo
  {year} {2013})}\BibitemShut {NoStop}%
\bibitem [{\citenamefont {di~Santo}\ \emph {et~al.}(2017)\citenamefont
  {di~Santo}, \citenamefont {Villegas}, \citenamefont {Burioni},\ and\
  \citenamefont {Mu{\~n}oz}}]{di2017simple}%
  \BibitemOpen
  \bibfield  {author} {\bibinfo {author} {\bibfnamefont {S.}~\bibnamefont
  {di~Santo}}, \bibinfo {author} {\bibfnamefont {P.}~\bibnamefont {Villegas}},
  \bibinfo {author} {\bibfnamefont {R.}~\bibnamefont {Burioni}}, \ and\
  \bibinfo {author} {\bibfnamefont {M.~A.}\ \bibnamefont {Mu{\~n}oz}},\
  }\href@noop {} {\bibfield  {journal} {\bibinfo  {journal} {Physical Review
  E}\ }\textbf {\bibinfo {volume} {95}},\ \bibinfo {pages} {032115} (\bibinfo
  {year} {2017})}\BibitemShut {NoStop}%
\bibitem [{\citenamefont {Priesemann}\ \emph {et~al.}(2014)\citenamefont
  {Priesemann}, \citenamefont {Wibral}, \citenamefont {Valderrama},
  \citenamefont {Pr{\"o}pper}, \citenamefont {Le~Van~Quyen}, \citenamefont
  {Geisel}, \citenamefont {Triesch}, \citenamefont {Nikoli{\'c}},\ and\
  \citenamefont {Munk}}]{priesemann2014spike}%
  \BibitemOpen
  \bibfield  {author} {\bibinfo {author} {\bibfnamefont {V.}~\bibnamefont
  {Priesemann}}, \bibinfo {author} {\bibfnamefont {M.}~\bibnamefont {Wibral}},
  \bibinfo {author} {\bibfnamefont {M.}~\bibnamefont {Valderrama}}, \bibinfo
  {author} {\bibfnamefont {R.}~\bibnamefont {Pr{\"o}pper}}, \bibinfo {author}
  {\bibfnamefont {M.}~\bibnamefont {Le~Van~Quyen}}, \bibinfo {author}
  {\bibfnamefont {T.}~\bibnamefont {Geisel}}, \bibinfo {author} {\bibfnamefont
  {J.}~\bibnamefont {Triesch}}, \bibinfo {author} {\bibfnamefont
  {D.}~\bibnamefont {Nikoli{\'c}}}, \ and\ \bibinfo {author} {\bibfnamefont
  {M.~H.}\ \bibnamefont {Munk}},\ }\href@noop {} {\bibfield  {journal}
  {\bibinfo  {journal} {Frontiers in systems neuroscience}\ }\textbf {\bibinfo
  {volume} {8}},\ \bibinfo {pages} {108} (\bibinfo {year} {2014})}\BibitemShut
  {NoStop}%
\bibitem [{\citenamefont {Ogata}(1981)}]{ogata1981lewis}%
  \BibitemOpen
  \bibfield  {author} {\bibinfo {author} {\bibfnamefont {Y.}~\bibnamefont
  {Ogata}},\ }\href@noop {} {\bibfield  {journal} {\bibinfo  {journal} {IEEE
  transactions on information theory}\ }\textbf {\bibinfo {volume} {27}},\
  \bibinfo {pages} {23} (\bibinfo {year} {1981})}\BibitemShut {NoStop}%
\bibitem [{\citenamefont {Rizoiu}\ \emph {et~al.}(2017)\citenamefont {Rizoiu},
  \citenamefont {Lee}, \citenamefont {Mishra},\ and\ \citenamefont
  {Xie}}]{rizoiu2017tutorial}%
  \BibitemOpen
  \bibfield  {author} {\bibinfo {author} {\bibfnamefont {M.-A.}\ \bibnamefont
  {Rizoiu}}, \bibinfo {author} {\bibfnamefont {Y.}~\bibnamefont {Lee}},
  \bibinfo {author} {\bibfnamefont {S.}~\bibnamefont {Mishra}}, \ and\ \bibinfo
  {author} {\bibfnamefont {L.}~\bibnamefont {Xie}},\ }\href@noop {} {\bibfield
  {journal} {\bibinfo  {journal} {arXiv preprint arXiv:1708.06401}\ } (\bibinfo
  {year} {2017})}\BibitemShut {NoStop}%
\bibitem [{\citenamefont {Dassios}\ and\ \citenamefont
  {Zhao}(2011)}]{dassios2011dynamic}%
  \BibitemOpen
  \bibfield  {author} {\bibinfo {author} {\bibfnamefont {A.}~\bibnamefont
  {Dassios}}\ and\ \bibinfo {author} {\bibfnamefont {H.}~\bibnamefont {Zhao}},\
  }\href@noop {} {\bibfield  {journal} {\bibinfo  {journal} {Advances in
  applied probability}\ }\textbf {\bibinfo {volume} {43}},\ \bibinfo {pages}
  {814} (\bibinfo {year} {2011})}\BibitemShut {NoStop}%
\bibitem [{\citenamefont {Colomer-de Sim{\'o}n}\ and\ \citenamefont
  {Bogu{\~n}{\'a}}(2014)}]{colomer2014double}%
  \BibitemOpen
  \bibfield  {author} {\bibinfo {author} {\bibfnamefont {P.}~\bibnamefont
  {Colomer-de Sim{\'o}n}}\ and\ \bibinfo {author} {\bibfnamefont
  {M.}~\bibnamefont {Bogu{\~n}{\'a}}},\ }\href@noop {} {\bibfield  {journal}
  {\bibinfo  {journal} {Physical Review X}\ }\textbf {\bibinfo {volume} {4}},\
  \bibinfo {pages} {041020} (\bibinfo {year} {2014})}\BibitemShut {NoStop}%
\bibitem [{\citenamefont {Bogun{\'a}}\ \emph {et~al.}(2004)\citenamefont
  {Bogun{\'a}}, \citenamefont {Pastor-Satorras},\ and\ \citenamefont
  {Vespignani}}]{boguna2004cut}%
  \BibitemOpen
  \bibfield  {author} {\bibinfo {author} {\bibfnamefont {M.}~\bibnamefont
  {Bogun{\'a}}}, \bibinfo {author} {\bibfnamefont {R.}~\bibnamefont
  {Pastor-Satorras}}, \ and\ \bibinfo {author} {\bibfnamefont {A.}~\bibnamefont
  {Vespignani}},\ }\href@noop {} {\bibfield  {journal} {\bibinfo  {journal}
  {The European Physical Journal B}\ }\textbf {\bibinfo {volume} {38}},\
  \bibinfo {pages} {205} (\bibinfo {year} {2004})}\BibitemShut {NoStop}%
\end{thebibliography}%

\newpage

\renewcommand{\theequation}{S\arabic{equation}}
\renewcommand{\thefigure}{S\arabic{figure}}

\onecolumngrid
%\appendix

\section*{Supplemental Material}

\subsection{Hawkes process: numerical simulations}
\label{sec:Hsimul}

\subsubsection{Exponential kernel}

The Markov property of the exponential kernel can be exploited to numerically simulate
the process in linear time by means of the algorithm developed in Ref.~\cite{dassios2013exact}. In the present case of unmarked Hawkes process the 
implementation is straightforward. Given the rate $\lambda(t_{k-1})$ right after the $(k-1)$-th event, the $k$-th inter-event time can be sampled as
\begin{enumerate}
    \item Set $F_1= -\log(u_1)/\mu, \text{ where } u_1 \sim \text{Unif}(0, 1)$.
    \item Set $G_2 = 1 + \log(u_2)/(\lambda(t_{k-1}) - \mu), \text{ where } u_2 \sim \text{Unif}(0, 1)$.
    \item Set $F_2 = 
    \begin{cases}
        \infty \, \text{ \quad \quad \quad if } G_2 \leq 0\\
        -\log(G_2) \text{ \, if } G_2 > 0
    \end{cases}$ .
    \item Set $x_k = \text{min}\{ F_1, F_2\}$.
    \item Set $\lambda(t_{k-1}+x_k) = (\lambda(t_{k-1})-\mu) e^{-x_k} + n + \mu$.
\end{enumerate}
\begin{comment}
The above algorithm can be implemented to simulate the process
for $\mu=0$ as well, with the unique caveat that the initial condition must be
set manually (while Eq.~(1) intrinsically has the initial condition
$\lambda = \mu$). In particular for $\mu=0$ $F_1=\infty$ so the run proceeds
until $F_2=\infty$ as well. This simulation scheme allows to fast sample $K$
events for $\mu>0$ and a random number of events for $\mu=0$. This random number has
the distribution of the size of a branching process, i.e., a power law distribution with
the exponent $3/2$. 
\end{comment}

%The Hawkes process with power law kernel is not Markovian and the above algorithm cannot
%be used. 
%The thinning algorithm by Ogata~\cite{ogata1981lewis}, on the other hand, can be used
%for any kernel as specified in the main text, but it is quadratic in the number of events.

\subsubsection{Power-law kernel}

To generate time series from an Hawkes process with power-law kernel we take advantage
of the thinning algorithm by Ogata~\cite{ogata1981lewis}.
The computational complexity of the algorithm grows quadratically with the number of events.
The thinning algorithm requires knowledge of the full history 
$\{t_1, t_2, ..., t_{k-1}\}$ of the process for the $k$-th inter-event to be sampled.
Specifically, given $\lambda(t_{k-1})$ and the set of previous event times, 
\begin{enumerate}
    \item Set $x = -\log(u_1)/\lambda(t_{k-1}), \text{ where } u_1 \sim \text{Unif}(0, 1)$.
    \item Compute $\lambda(t_{k-1}+x)$ and compute the ratio 
    $\ell = \lambda(t_{k-1}+x)/\lambda(t_{k-1})$.
    \item Sample $u_2 \sim \text{Unif}(0, 1)$.
    \item If $u_2 \leq \ell$, set $t_k = t_{k-1}+x$. Note that 
    $\lambda(t_{k}) = \lambda(t_{k-1}+x) + n (\gamma-1)$, where the first term on the right hand side has been computed at step 2.
\end{enumerate}
Note that, for efficiency reasons, time can be updated even if the event is rejected (i.e.
$u_2 > \ell$) and $\lambda(t_{k-1}+x)$ can be used in step 1 instead of
$\lambda(t_{k-1})$~\cite{rizoiu2017tutorial}. 
%Given an initial condition
%$\lambda(0)$, this algorithm can be implemented even for $\mu=0$. However, since
%\lambda$ may reach extremely small values and produce
%exceedingly large inter-event times, a reasonable implementation of this algorithm
%for $\mu=0$ requires an exit condition based on a minimal (maximal) value of $\lambda$ ($x$).

\subsubsection{Leveraging the equivalence with the branching process to generating time series for the Hawkes process}

As stated in the main text, a statistically equivalent way to produce time series with
the rate of Eq.~(1) is to generate a branching tree with the Poisson 
offspring distribution $P(o; n)$ and then to assign an event time to each node. The 
inter-event time $x$ between successive generation's nodes is a random variable with
distribution $\phi(x)$. 
This type of simulations is particularly useful to explore the large-scale properties of the process with the power-law kernel, in that the quadratic scaling
of the thinning algorithm with $K$ prevents large time series to be generated.

\subsection{Critical Hawkes process: average rate}
\label{sec:H_averageRate}

% Moments and fluctuations

Theorem 3.6 and Corollary 3.5 in Ref.~\cite{dassios2011dynamic} can be reformulated for
the rate of Eq.~(1) in the case of exponential kernel. The rate in Ref.~\cite{dassios2011dynamic} takes
the form of Eq.~(1) by simply setting, in Ref.~\cite{dassios2011dynamic} notation, $a=\lambda_0=\mu$, $\delta=1$, 
$Y_i=0$ and $Z_i = n$ for all $i$. Then Theorem 3.6 and Corollary 3.5 take the form
\begin{equation}
    \begin{split}
    & \av{\lambda(t)} = \mu(1+t) \\
    & \av{\lambda^2(t)} = \mu^2 +(2\mu + n^2)(\mu t + \mu t^2 / 2) 
    \end{split} \, 
    \label{eq:rate_moments}
\end{equation}
respectively for $n=1$. 
In Eq.~(\ref{eq:rate_moments}), $\langle \cdot \rangle$ indicates average value
over an infinite number of realizations of the Hawkes process. 
It follows that 
\begin{equation}
  \frac{\av{\lambda^2(t)}-\av{\lambda(t)}^2}{\av{\lambda(t)}^2} =
  \frac{1}{2\mu}
  \label{eq:flucts}
\end{equation}
for large time. Thus the process experiences fluctuations that are much smaller than 
the average rate for $\mu \gg 1$ and vice versa. By noting that the number of events at time $t$, namely $k(t)$, 
obeys $dk(t)/dt = \lambda(t)$,
%is given by
%$\dot{k}(t) = \lambda(t)$, 
it is easy to see that 
$\av{k(t)} = \mu\left(t + t^2/2\right)$. It follows that for $t \ll 1$ the 
Hawkes process behaves as a Poisson process, being $\av{\lambda} \approx \mu$ and 
$\av{k}(t) \approx \mu t$, while for large time we have that $\av{\lambda} \approx \mu t$ and
$\av{k}(t) \approx \mu t^2/2$. Finally, inverting the relation between time and number of events,
the average rate is
\begin{equation}
    \av{\lambda(k)} = \mu + \sqrt{2 \mu k} \, .
    \label{eq:av_rate}
\end{equation}

\begin{figure}[!htb]
\begin{center}
\includegraphics[width=0.95\textwidth]{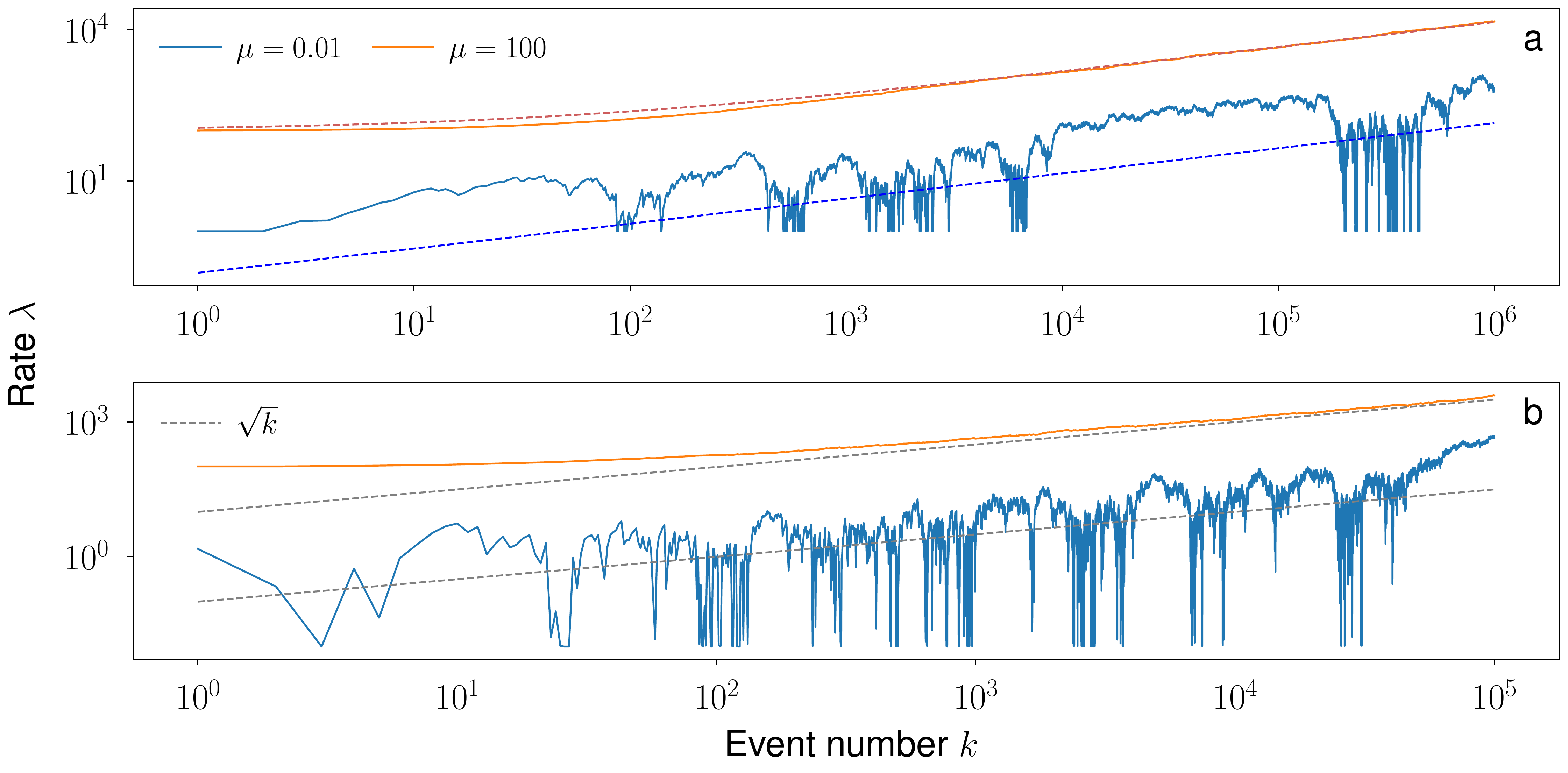}
\end{center}
\caption{Individual realizations of the critical Hawkes process. (a) Exponential kernel.
Two realizations of the process, one for $\mu=0.01$ and one for $\mu=100$. Each realization is
compared with the average rate in Eq.~\ref{eq:rate_moments}, shown
as a dashed line. (b) Same as before but using a power law kernel with $\gamma=2.5$. The dashed
line here simply scales as $\sqrt{k}$.}
\label{fig:SM1}
\end{figure}

The equations above 
allow to understand 
the physical interpretation of the phase diagram of Figure~1 of the main paper. In Figure~\ref{fig:SM1}a, the average
rate is shown and compared to individual realizations of the process. For $\mu \ll 1$, the individual
realization is characterized by a sequence of bursts during which the rate grows proportionally to the
square root of the event number, in agreement with Eq.~(\ref{eq:av_rate}). A burst ends 
when a fluctuation of the rate is large enough for it to drop much below its average value. 
%These 
Strong fluctuations 
%are due to
correspond to
inter-event times much larger than $1$. %and are expected by virtue of
They are expected to occur as predicted by Eq.~(\ref{eq:flucts}). 
%Such 
Large inter-event times make all the individual terms in the kernel small, 
%and the rate drops 
with the rate dropping
to its minimal value, i.e., $\lambda=\mu$. 
Then, a new event occurring after
an inter-event time of the order of
$\mu^{-1}$ 
%excites 
gives rise to
a new burst of the self-exciting process. When instead $\mu \gg 1$, 
%the 
fluctuations are 
expected to be smaller than the average value. 
%As explained in the main text, this 
%is due to the fact that 
The temporal scales of the spontaneous activity %(Poisson process with rate $\mu$) 
and of
the self-exciting process 
are 
%to be 
comparable, so that spontaneously generated events reinforce the 
burst instead of interrupting it. 

The quadratic scaling 
$\av{k(t)} \approx \mu t^2/2$
observed at large times is the signature of 
the branching process underlying the evolution of $\lambda$. The number of events grows 
%indeed 
as the square of time, in the same way as the size of a branching process grows as the square
of the number of generations. This 
fact
can also be seen by noting that $\lambda \sim \sqrt{k}$ and that
the typical inter-event time is of the order of $1/\lambda$. 
We are not aware of explicit formulas for the average rate in the case 
of the power-law kernel.
However, it is not surprising to see, in numerical simulations, 
%that numerical simulation show 
%the growth of 
$\lambda$ 
growing
with the square root of $k$ for this kernel as well
%, as shown in 
(Figure~\ref{fig:SM1}b).

\begin{comment}
 In particular, the map from Hawkes process to branching process
relies on spontaneous events to act as seeds of branching trees. For $\mu=0$ PROBABLY NOT NEEDED: 
descrizione delle singole realizzazioni per mu=0. Concetto di mu_eff. 
\end{comment}

%{\bf Filippo's recursive equation (?). Does it work for $n\neq 1$?}

\subsection{Finite-size scaling analysis}
\label{sec:FSS}

The percolation strength is defined as the average
size of the largest cluster, $S_M$, normalized to $K$:
\begin{equation}
    P_{\infty} = \frac{\av{S_M}}{K} \, .
    \label{eq:Pinf}
\end{equation}
The susceptibility associated to the order parameter can be defined as
\begin{equation}
    \chi = \frac{\av{S_{M}^2} - \av{S_{M}}^2}{\av{S_{M}}} \, .
    \label{eq:chi}
\end{equation}
Averages are meant over 
an infinite number of realizations of the Hawkes process.
%realizations.

The cluster number $n_S$ is usually defined as the number of clusters
of size $S$ per unit lattice site~\cite{stauffer2018introduction}.
The same definition can be used for the cluster duration $n_T$.
In percolation theory, the scaling ansatz for the cluster number is
formulated as
\begin{equation}
    \begin{split}
    &   n_S = S^{-\tau} \mathscr{G}_{S} (z) \\
    &   z = (p_c-p)S^{\sigma} \, ,
    \end{split} 
    \label{eq:size_ansatz}
\end{equation}
where $p$ is the concentration (i.e., the bond occupation probability) and
$p_c$ is the critical point of the model. Homogeneous percolation
in one dimension has $p_c=1$.
The scaling ansatz is assumed to be valid in the vicinity
of the critical point and for sizes $S \gg 1$~\cite{stauffer2018introduction}. 
Despite the scaling function
is, in principle, unknown, it is expected to decay fast after a crossover
whose typical scale is
\begin{equation}
    S_c \propto |p-p_c|^{-1/\sigma} \, .
    \label{eq:size_crossover}
\end{equation}
Equations~(\ref{eq:size_ansatz},~\ref{eq:size_crossover}) define the exponents $\tau$ and $\sigma$, respectively
quantifying the decay of the cluster number and the divergence of the 
crossover scale. %These are the only
%two relevant exponents: 
All the other critical exponents can be computed via scaling
relations involving $\tau$ and $\sigma$.
Cluster duration, on a lattice, is the same observable as the cluster size.
On disordered topologies, we assume that a scaling
form analogous to Eq.~(\ref{eq:size_ansatz}) exists for the 
cluster duration, with $n_T$
characterized by its own exponents $\alpha$ and $\sigma_T$ and by
its own scaling function $\mathscr{G}_T$.

%From the scaling of the cluster number, i.e., Eq.~(\ref{eq:size_ansatz}), 
%it is easy to derive
%the behavior of the order parameter in the vicinity of the critical point.
Given the scaling of Eq.~(\ref{eq:size_ansatz}), 
the order parameter is expected to grow from zero to a positive value as a power
law with exponent $\beta$. The critical exponent $\beta$ can be determined using the knowledge of
$\tau$ and $\sigma$, according to
\begin{equation}
    \begin{split}
    &    P_{\infty} \propto (p-p_c)^{\beta} \text{ for } p \rightarrow p_c^+ \, , \\
    &    \beta = \frac{\tau-2}{\sigma} \, .
    \end{split}
    \label{eq:beta}
\end{equation}
In the vicinity of the critical point the susceptibility~(\ref{eq:chi})
diverges as a power law
\begin{equation}
    \begin{split}
    &   \chi \propto |p-p_c|^{-(\gamma+\beta)} \\
    &   \gamma = \frac{3-\tau}{\sigma} \, .
    \end{split}
    \label{eq:gamma}
\end{equation}
The presence of the exponent $\beta$ in the above
expression is direct a consequence of the 
definition of Eq.~(\ref{eq:chi})~\cite{colomer2014double}. 
%If the normalization
%$\av{S_{M}}$ is replaced by the more common system size, the susceptibility
%diverges with exponent $\gamma$. However, if the exponent $\beta>0$,
%then the exponent of $\chi$ is larger than simply $\gamma$, making the 
%numerical measure easier.

Critical exponents can be evaluated by means of Finite-Size Scaling (FSS) analysis.
The key concept underlying FSS is the existence of a unique
length scale characterizing the behaviour of an infinite system.
This is the correlation length $\xi$, which represents
the typical linear size of the largest (but finite) cluster.
The correlation length is in fact defined as the typical scale of the 
correlation function, which has the exponential form $g(r)=e^{-r/\xi}$ for
$p<p_c$. The correlation function, in turn, is defined as the probability that a site
at distance $r$ from an active site belongs to the same cluster.
The correlation length diverges as $|p-p_c|^{-\nu}$ in the 
vicinity of the critical point, indicating that the largest cluster
spans the whole lattice. This divergence, however, is de facto bounded
by $K$ in finite systems, with $K$ the linear
size of the system. Then in any finite system, at any
$p$ value, there are two possibilities: $p$ is close enough to $p_c$, so
that $\xi \gg K$, or $p$ is far enough from $p_c$ and $\xi \ll K$.

A second crucial property is the nature of the
critical point: $p_c$ is a thermodynamic quantity. Critical behavior in finite
systems is observed at a different value of the 
concentration $p$, namely $p^*$. The pseudo (size-dependent) critical 
point 
%can be estimated in numerous different ways and any reasonable
%definition of $p^*$ 
%is expected to 
scales with the system size according to
\begin{equation}
p_c - p^*(K) \sim K^{-1/\nu} \, .
\label{eq:nu}
\end{equation}
The above equation quantifies the convergence of $p^*$ toward $p_c$, 
and allows us to estimate the exponent $\nu$. In the vicinity of 
$p^*$, we further have $\xi \gg K$, thus allowing us to measure 
the exponents $\beta$ and $\gamma$. %as well: 
The scaling 
$P_{\infty} \sim (p-p_c)^{\beta}$ can be replaced, by virtue of 
$\xi \sim (p-p_c)^{-\nu}$, with $P_{\infty} \sim \xi^{-\beta/\nu}$. Further, if the system is close enough to criticality, $\xi$ is bounded by $K$ and
the expected scaling is $P_{\infty} \sim K^{-\beta/\nu}$.
Analogous reasoning leads to the scaling of
the susceptibility $\chi \sim  K^{-(\gamma+\beta)/\nu}$.
We recall that ordinary percolation in one dimension is characterized by the exponents
$\tau=2$, $\nu=\sigma=\gamma=1$ and $\beta=0$, stemming for the discontinuous nature of the transition.

\subsection{Percolation theory of the Poisson process}
\label{sec:Poisson}

The Hawkes process for $n=0$ is a Poisson process with rate $\mu$. 
Event times are independent, and the inter-event time distribution is exponential with 
rate $\mu$. It follows that the probability that two consecutive events belong to the same cluster is
$p = \mu \int_0^{\Delta} e^{-\mu x} \, dx = 1 - e^{-\mu \Delta}$. We used the symbol 
$p$ on purpose to stress the analogy of this quantity with the bond occupation probability in 
ordinary percolation. 
%Note that $p \rightarrow 1$ for $\Delta \rightarrow \infty$.

The percolation probability in a finite system can be computed as follows. 
A time series with $K$ events percolates if $\Delta$ is greater
than the largest inter-event time in the time series. We can estimate the effective
threshold $\Delta^*$ as the average (over realizations)
of the largest inter-event time observed by drawing $K$ samples from
the inter-event time distribution. 
%This quantity is the solution w.r.t. $\Delta$ of the equation
We can therefore solve the equation
\begin{equation*}
K \int_{\Delta^*}^{\infty} P(x) \, dx = 1 \, ,
\end{equation*}
%leading to 
obtaining
$\Delta^*(K) = \mu^{-1} \log(K)$.
%, where the subscript $P$ stands for the Poisson type 
%of transition. 

In summary, the percolation problem associated to time series generated by a homogeneous 
Poisson process is a
standard bond percolation problem in one dimension. 
In particular, all critical exponents are identical to those of the one-dimensional 
percolation, as well as the cluster number $n_S$ is expected to behave as predicted by the
exact solution in Ref.~\cite{stauffer1978critical}. The cluster duration $n_T$
can be also understood by nothing that, as the duration is the sum of the inter-event times
composing the cluster and as the inter-event time distribution is exponential, the duration of a
cluster with size $S$ is, on average, $\av{T} = \av{x} S = S \mu^{-1}$. The cluster duration
$n_T$ can be computed by noting that the total number of cluster with duration $T$ is 
$N_T = \mu N_s = \mu K n_s$ and that the cluster duration must be normalized to the length
of the system, $L = \av{x} K = \mu^{-1} K$, instead of its size $K$. In summary, one has
\begin{equation}
    \begin{split}
    & n_S = \frac{N_s}{K} = S^{-2} \mathscr{G}[S(p_c - p)] \\
    & n_T = \frac{N_T}{L} = T^{-2} \mathscr{G}[\mu T (p_c - p)] \\
    & \mathscr{G}(x) = x^2 e^{-x} \, .
     \end{split}
\end{equation}
Note also that the scaling function $\mathscr{G}_S = \mathscr{G}_T$,
so we use the symbol $\mathscr{G}$.

\begin{figure}[!htb]
	\begin{center}
	\includegraphics[width = 0.95\textwidth]{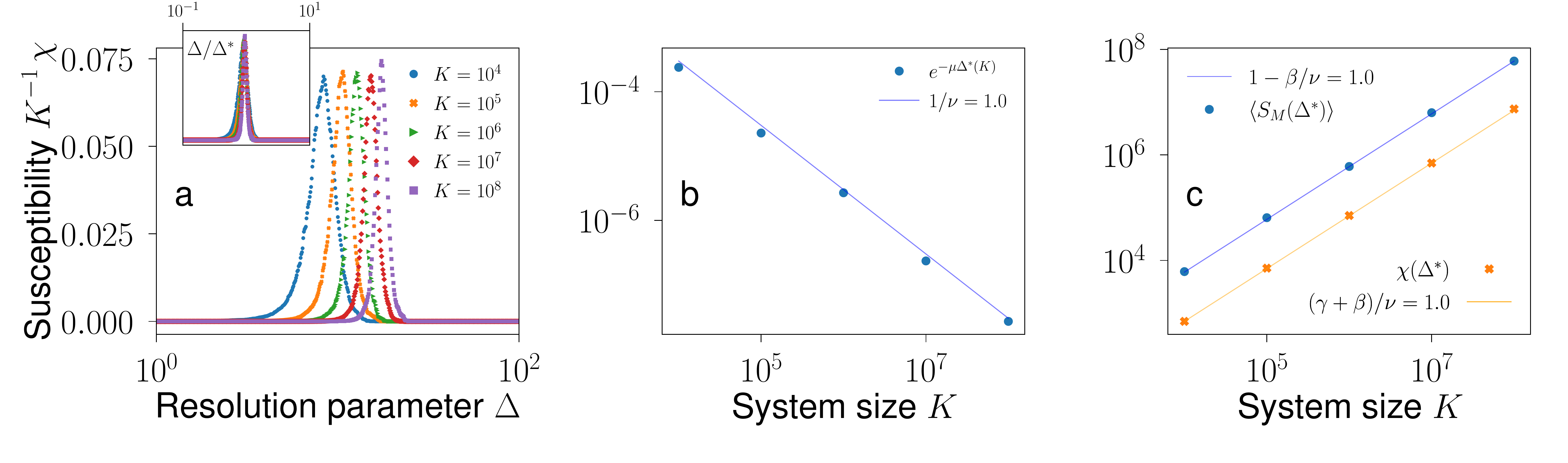}
	\end{center}
\caption{Finite-size scaling analysis of percolation in Poisson process. We use here $\mu=1$.
(a) Susceptibility [i.e., Eq.~(\ref{eq:chi})]
divided by 
$K$, as a function of the resolution parameter $\Delta$. The inset shows the same data
as of the main panel but the resolution parameter is rescaled as $\Delta/\Delta^*$.
(b) Convergence of the effective critical point to the critical point. We measure $\Delta^*(K)$ as the value of the resolution parameter where we observe the peak of the susceptibility. 
(c) Scaling of the peak of susceptibility (orange squares) and of the average size of the
largest cluster (blue circles) at the effective resolution $\Delta^*$.}
\label{fig:POIS_FSS}
\end{figure}

Figure~\ref{fig:POIS_FSS} shows the result of the FSS analysis for the Poisson process.
We measure 
%Measuring empirically 
the 
effective critical point as the value of the temporal resolution $\Delta$ 
%at which 
where
the susceptibility peaks. We use %allows to estimate 
$p_c - p^* = e^{-\mu \Delta^*}$
and confirm the scaling $K^{-1}$,  i.e., $\nu=1$. 
\changejan{To correctly determine the value of $\nu$, the scaling of the threshold with the system size should be performed by measuring system size in units of time and not of events. We note, however, that for the homogeneous Poisson process the number of events grows linearly with time, i.e., $k(t) \sim t$, thus measuring $\nu$ with respect to time or number of events does not change its value.}
The exponents $\gamma$ and $\beta$, 
measured from the value of $\chi$ and of $\av{S_M}$ at 
%the empirical
$\Delta^*$, 
are also consistent with those
of the ordinary one-dimensional percolation.
%, confirm the analogy between percolation of time series the Poisson process and
%ordinary one-dimensional percolation.

\begin{figure}[!htb]
	\begin{center}
	\includegraphics[width = 0.95\textwidth]{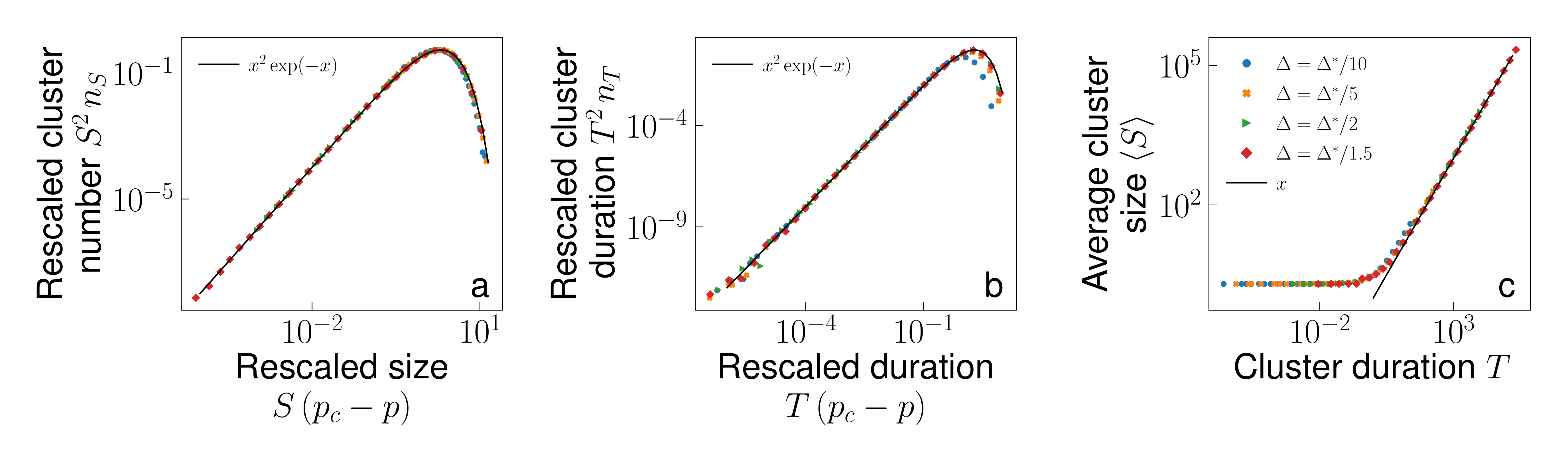}
	\end{center}
\caption{Cluster statistics of the Poisson process. We set $\mu = 1$. Each curve is estimated by 
considering $C=10^6$ clusters. The system size is $K=10^6$.
(a) Cluster number $n_S$, rescaled by $S^2$, as a function of
$S(p_c-p)$, for different values of $p$ and $p_c=1$. Here, $p = 1 - e^{-\mu \Delta}$.
The solid black line is the 
scaling function $\mathscr{G} (x)=x^2e^{-x}$~\cite{stauffer1978critical}
(b) Cluster duration $n_T$, 
rescaled by $T^2$, as a function of $T(p_c-p)$.
The solid black line is the same as in panel a.
(c) Average size of clusters as a function of their duration.
The black line indicates linear scaling.}
\label{fig:POIS_clusts}
\end{figure}

Also, we confirm the value of the critical exponents $\tau=\alpha=2$ and 
$\sigma=\sigma_T=1$ by studying the collapse of the cluster number on the scaling function $\mathscr{G}$ (Figure~\ref{fig:POIS_clusts}).

\begin{figure}[!htb]
	\begin{center}
	\includegraphics[width = 0.95\textwidth]{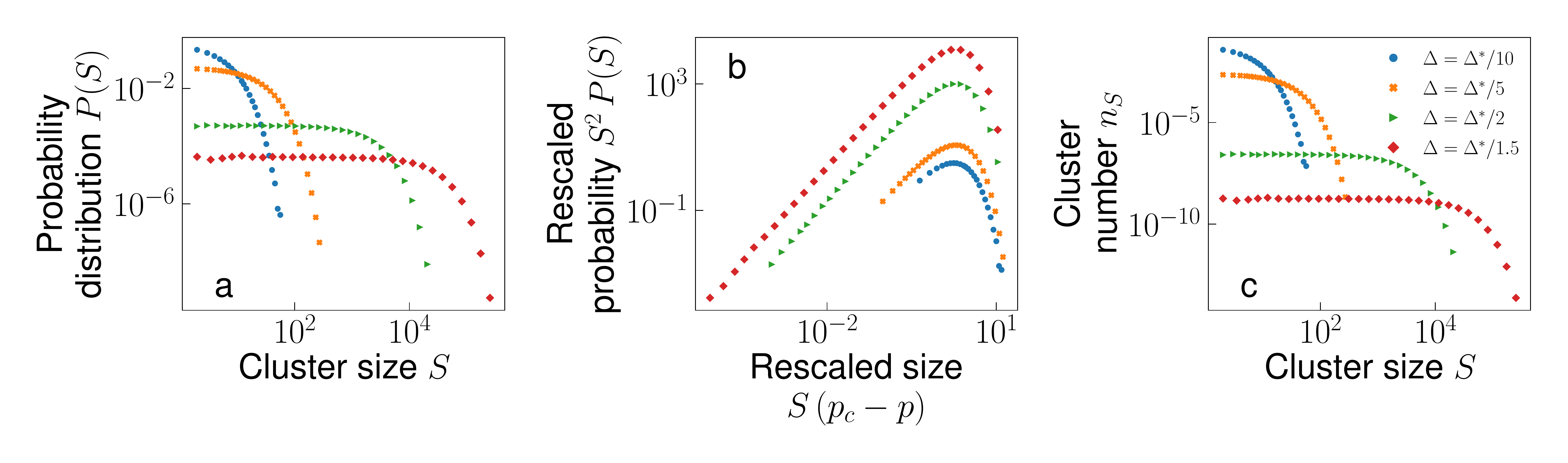}
	\end{center}
\caption{Cluster statistics of the Poisson process. 
We set $\mu = 1$. Each curve is estimated by 
considering $C=10^6$ clusters. The system size is $K=10^6$.
(a) Probability distribution $P(S)$ of the cluster size for different values of $\Delta$.
(b) Same as in panel a, but the distribution is rescaled by $S^2$ and the ordinate is rescaled by $(p_c-p)$, with $p = 1 e^{-\mu \Delta}$ and $p_c=1$.
(c) Cluster number $n_S$ for the same data as in panels a, b. 
}
\label{fig:POIS_PDF}
\end{figure}

We stress that the value of the critical exponent $\tau$ cannot be deduced immediately by looking at the distribution $P(S)$
at criticality (Figure~\ref{fig:POIS_PDF}). 
%Both these quantities measure the frequency of clusters of size $S$, but they differ
%in their normalization and this prevents any scaling to be observed from $P(S)$, as can be
%seen from Figure~\ref{fig:POIS_PDF}. However, the rescaled probability distribution does not collapses on the
%scaling function ${G}_{S} (x)$ and no scaling is observed.

\subsection{Percolation theory of the critical Hawkes process}
\label{sec:Hpercolation}

\begin{figure}[!htb]
	\begin{center}
	\includegraphics[width = 0.95\textwidth]{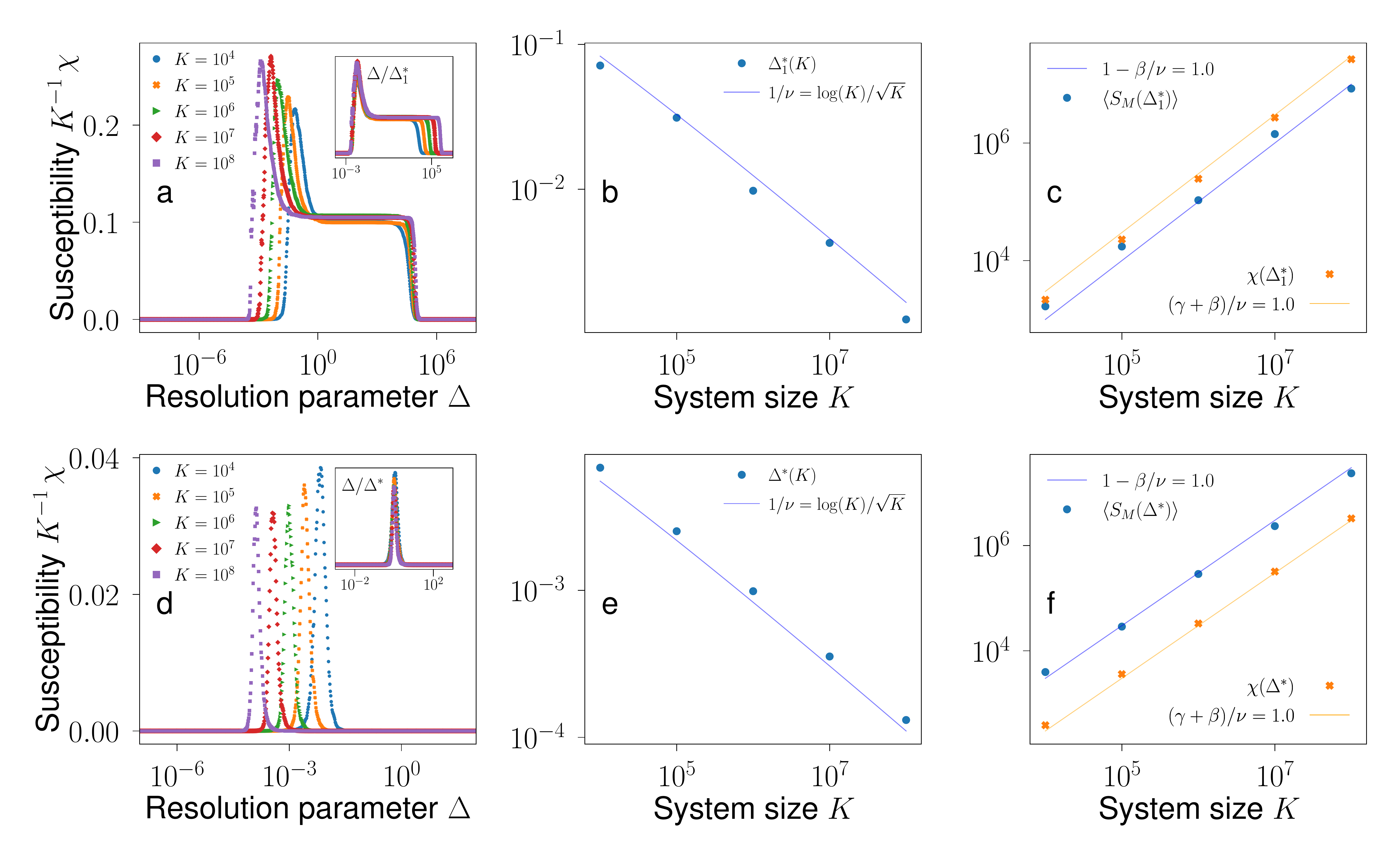}
	\end{center}
\caption{Finite-size scaling analysis of percolation in the Hawkes process. In panels a, b, and c, we use $\mu=10^{-4}$ . (a) Susceptibility [i.e., Eq.~(\ref{eq:chi})] divided by $K$ as a function of the resolution parameter $\Delta$. The inset shows the same data as in the main panel bu as a function of the rescaled variable $\Delta/\Delta_1^*$. (b) Scaling of the position
of the peak of the susceptibility for different system size. (c) Scaling of the maximum of the 
susceptibility and of the average size of the largest cluster at the resolution $\Delta^*_1$.
(d), (e), (f) Same as in panels a, b, and c, respectively, but for $\mu=10^2$.}
\label{fig:Hawkes_FSS}
\end{figure}

We can fully describe the
percolation transition of the Hawkes process %can be fully understood 
in terms
of the percolation transition of the 
Poisson process. 
The only caveat is accounting for the fact that process is not stationary at criticality~\cite{hawkes1974cluster}.
To this end, we simply assume that the process is a Poisson process
with rate dependent on the number
of events as in Eq.~(\ref{eq:av_rate}). The effective critical point 
%of the Poisson process is $\log(K)/\mu = \log(K)/\lambda$ so that the natural
%extension of this result to a non stationary process is to postulate
is
\[
\Delta_1^* = \frac{\log(K)}{\av{\lambda(K)}} = \frac{\log(K)}{\mu+\sqrt{2 K \mu}} \, .
\]
The above expression implies that, in the thermodynamic limit, the critical
point of the model is $\Delta_c = 0$ and that $\Delta_1^*$ approaches $\Delta_c$
with the exponent $\nu=2$ and a logarithmic correction. In Figure~1 of the main text, we have shown that the expression allows us to 
properly rescale the order parameter into a unique scaling function. The result is confirmed in 
Figure~\ref{fig:Hawkes_FSS}. 
\changejan{The finding is compatible with the one valid for the homogeneous Poisson process. If the scaling of the threshold with system size is measured in units of time, we then recover $\nu=1$. In fact, the number of events scales quadratically with time, i.e., $k(t) \sim t^2$.
}
There, the susceptibility associated to the order parameter
collapses, if properly rescaled, 
on a unique curve,  %under the same rescaling 
irrespective of the $\mu$ value under consideration.
The empirical measure of the effective critical point, as the value of $\Delta$ where the
susceptibility peaks, further shows the scaling $\Delta_1^*(K) \sim \log(K)/K^{1/2}$. The FSS analysis
is completed by showing that the exponent $\beta=0$, as expected for a discontinuous transition, and
that the exponent $\gamma=\nu$. It follows, from respectively Eqs.~(\ref{eq:beta}) and~(\ref{eq:gamma}), that
$\tau=2$ and $\sigma=2$.

\begin{figure}[!htb]
	\begin{center}
	\includegraphics[width = 0.95\textwidth]{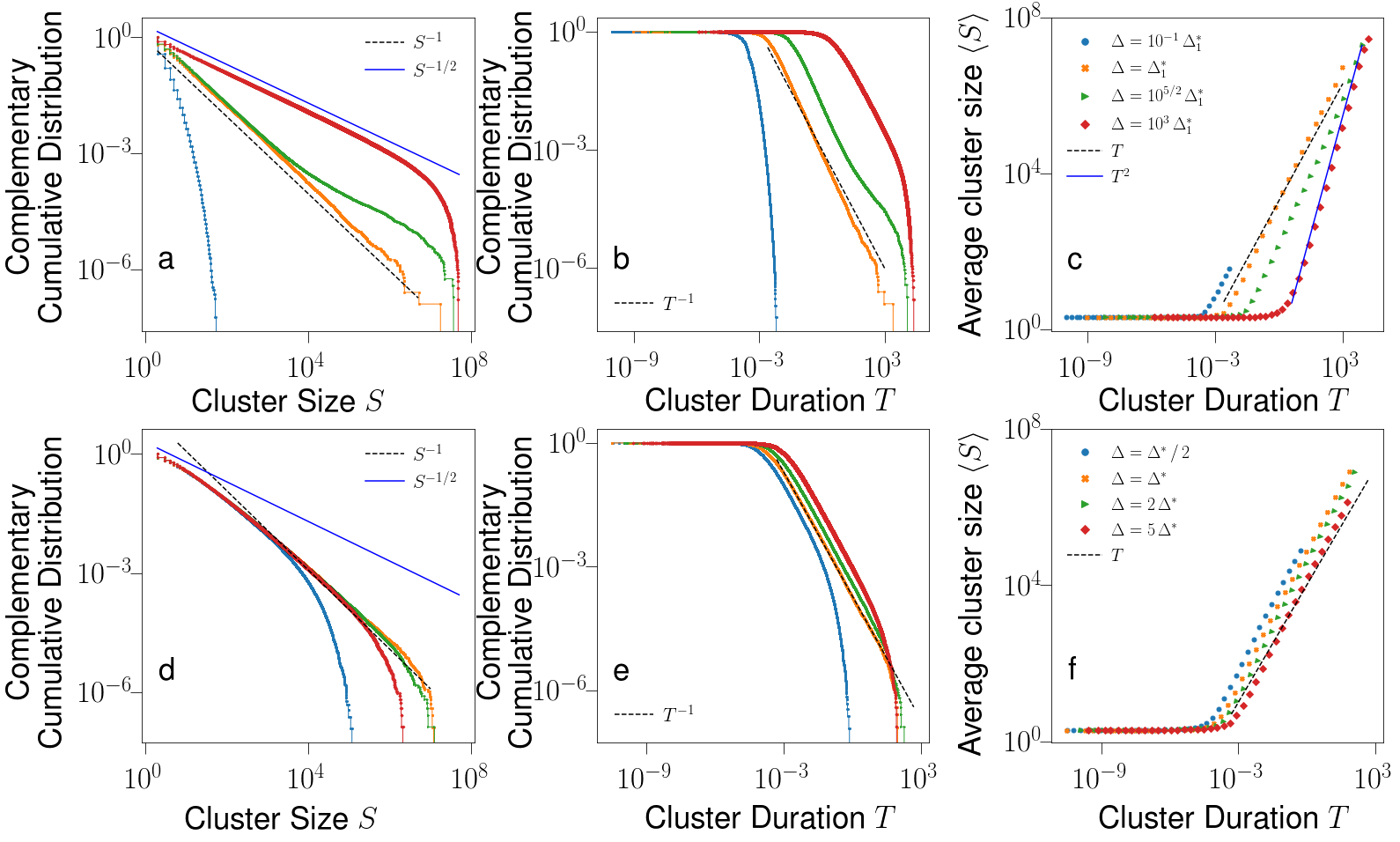}
	\end{center}
\caption{
Critical  percolation  properties  of  self-exciting  temporal processes. We consider  the rate of
Eq.~(1) of the main text with exponential kernel function and $n=1$.  System size is fixed at $K=10^8$. 
Histograms are obtained by considering $C=10^7$ clusters per configuration. We use $\mu=10^{-4}$ in
panels a, b, and c. (a) Complementary Cumulative Distribution of the cluster size at 
different resolution $\Delta$. (b)  Distribution  of  the  cluster  duration for the same data as in
panel a. (c) Average size of clusters as a function of their duration for the same data as in panel
(a). (d), (e) and (f) Same as in panels a, b, and c, respectively, but for $\mu=10^2$.}
\label{fig:Hawkes_clusts}
\end{figure}

The distribution of the cluster sizes and durations have been shown in Figure~2 of the main text. 
Figure~\ref{fig:Hawkes_clusts} displays
the complementary cumulative distribution for the same data in Figure~1 of the main text, confirming the results shown in the main text. 
Also, we display the linear scaling
of the average size of clusters with given duration, as
expected from the fact that both critical exponents $\tau = \alpha = 2$.
For $\mu \ll 1$ and $\Delta$ large enough, the cluster size distribution shows the scaling $\tau=3/2$. For this choice of the parameters, we observe also a quadratic relation between $S$ and $T$.

\begin{figure*}[!htb]
\includegraphics[width=0.85\textwidth]{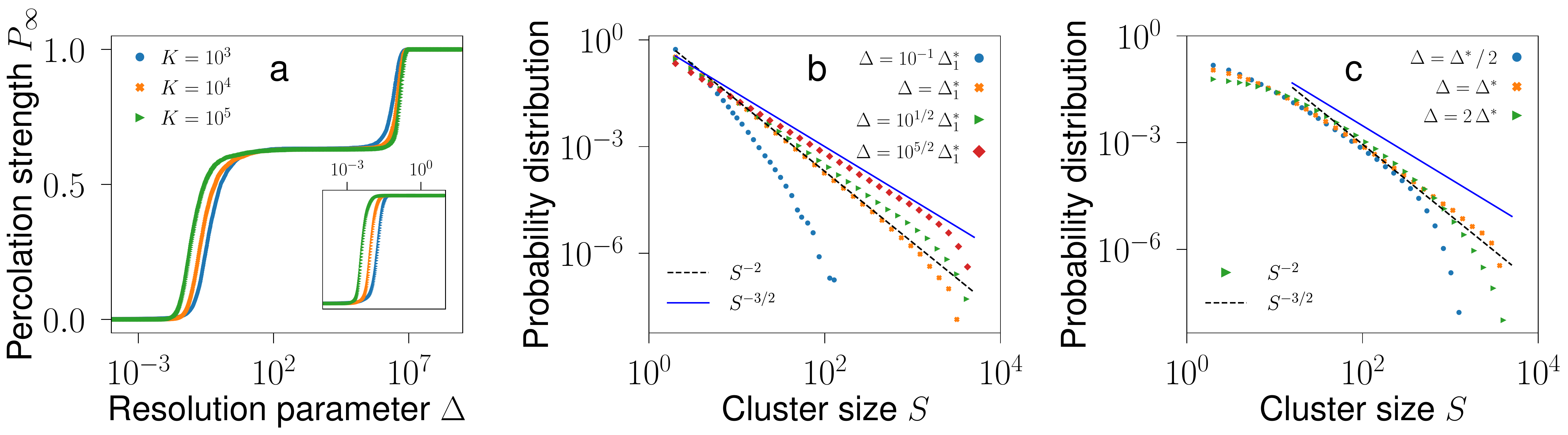}
\caption{Percolation transition in power-law self-exciting temporal processes. 
We consider temporal processes generated by the rate of Eq.~(1) of the main text with a 
power-law kernel and parameter $n=1$. The exponent of the power-law kernel is 
$\gamma = 2.5$. Results are obtained by considering $R = 10^3$ realizations of 
the process. (a) Percolation strength as a function of the resolution parameter. 
We display results for different system sizes $K$. In the main panel, we set 
$\mu=10^{-6}$. In the inset, results are obtained for $\mu=10^2$. (b) Distribution 
of the cluster size at criticality  for  $\mu=10^{-6}$. System size here is 
$K = 10^4$. (c) Same as in panel b but for $\mu=10^2$.}
\label{fig:HPL_clusters}
\end{figure*}

As stated in the main text, we further consider the power law kernel
$\phi(x) = (\gamma-1)(1+x)^{-\gamma}$ for $\gamma>2$, so that $\phi(x)$ 
has a finite mean. The simulation requires the use of the thinning algorithm, so
that the system size $K$ must be kept small. We show the percolation
phase diagram and the cluster size distribution for both $\mu \ll 1$ and
$\mu \gg 1$ in Fig.~\ref{fig:HPL_clusters}. The phenomenology is clearly 
analogous, with a double transition for $\mu \ll 1$ and a single transition
for $\mu \gg 1$. Further, as Figure~\ref{fig:SM1} shows, the rate grows again 
as the square root of the number of events, as expected for a branching process.
It follows that the critical point $\Delta_1^*$ for the power-law kernel scales 
again as $K^{-1/2}$ as can be verified by rescaling the data in Figure~\ref{fig:HPL_clusters}a. Overall, Figure~\ref{fig:SM1} and Figure~\ref{fig:HPL_clusters} show that the results presented in the main text for the exponential kernel remain valid for the power law kernel which, in turn,
suggests that the results are valid for any kernel with finite mean. Large-scale
simulations of the non-Markovian Hawkes process can be performed by exploiting the
map of the process with the branching process (BP), which is done in the next section.

\subsection{The temporal branching process}
\label{sec:BP}

\begin{figure}[!htb]
	\begin{center}
	\includegraphics[width = 0.95\textwidth]{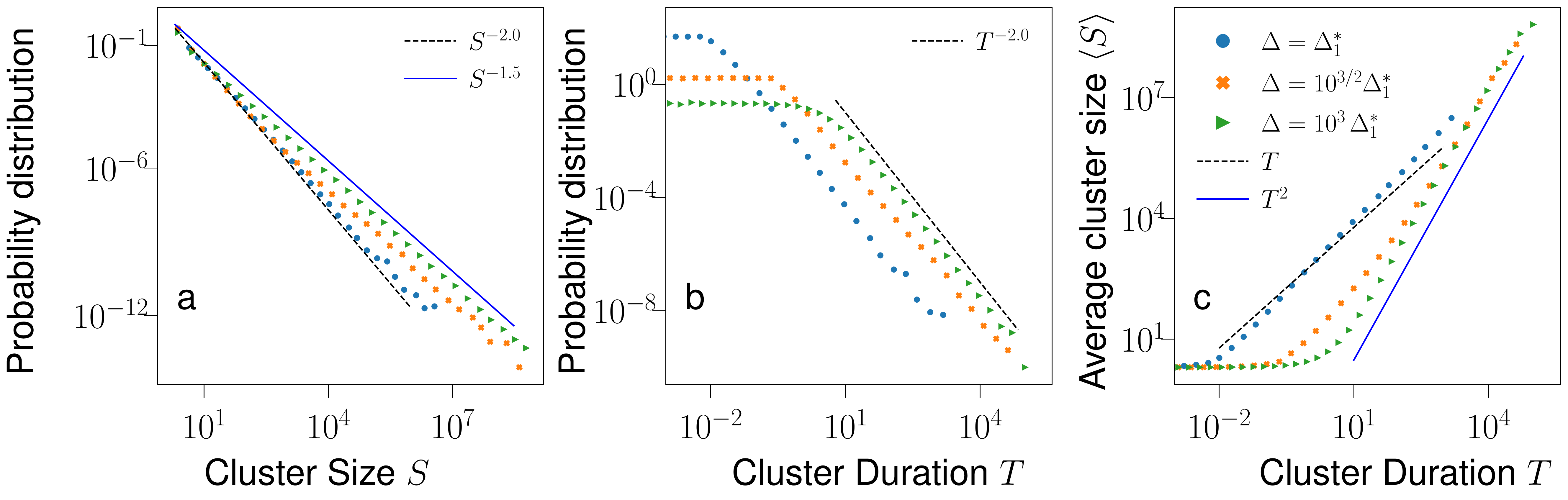}
	\end{center}
\caption{Critical  percolation  properties  of time series generated from a branching tree with 
bonds weighted by inter-event times distributed as  $\phi(x)$. We use here $\phi(x)=e^{-x}$.
Histograms are obtained by considering $C=10^6$ clusters per configuration.
(a) Distribution of the cluster size at 
different resolution $\Delta$. (b)  Distribution  of  the  cluster  duration for the same data as in panel a. (c) Average size of clusters as a function of their duration for the same data as in panel a.}
\label{fig:NHPP_exp}
\end{figure}

\begin{figure}[!htb]
	\begin{center}
	\includegraphics[width = 0.95\textwidth]{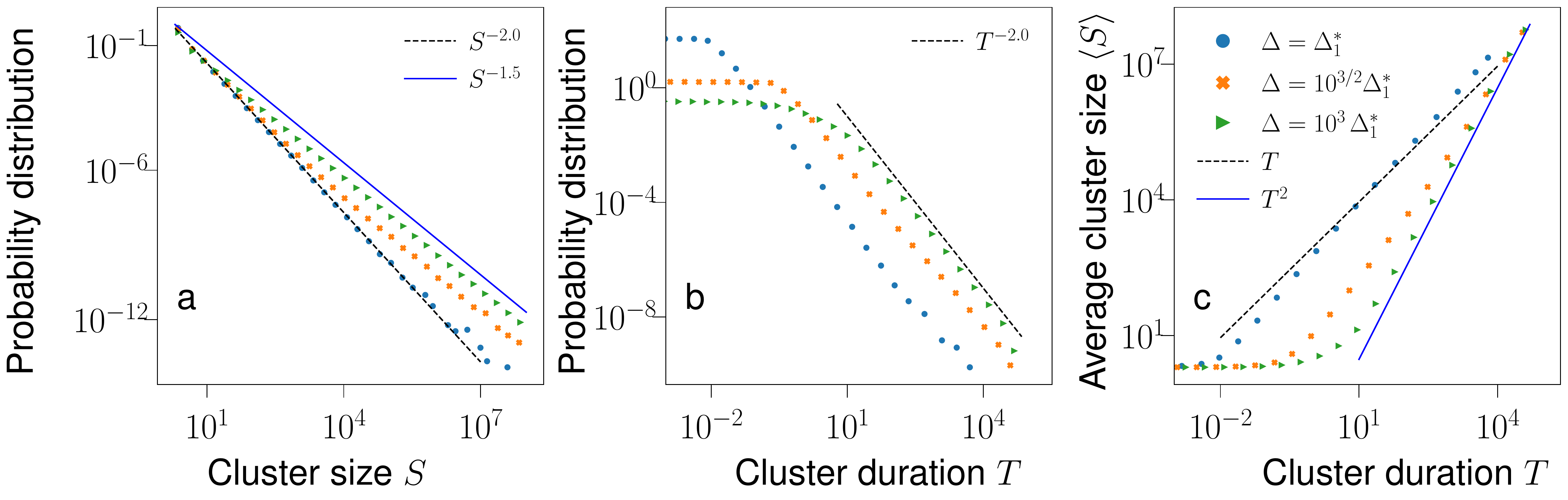}
	\end{center}
\caption{Critical  percolation  properties  of time series generated from a branching tree with 
bonds weighted by inter-event times distributed as  $\phi(x)$. We use here the power-law kernel $\phi(x) = (\gamma-1) (1+x)^{-\gamma}$ with $\gamma=2.5$.
Histograms are obtained by considering $C=10^6$ clusters per configuration. 
(a) Distribution of the cluster size at 
different resolution $\Delta$. (b)  Distribution  of  the  cluster  duration for the same data as in panel a. (c) Average size of clusters as a function of their duration for the same data as in panel a.}
\label{fig:NHPP_pl}
\end{figure}

To numerically demonstrate the statistical equivalence between the Hawkes process and the branching process, in Figures~\ref{fig:NHPP_exp} and \ref{fig:NHPP_pl} we show results valid for time series generated using a branching process with inter-event times respectively sampled from exponential and power-law distributions.

\subsection{Crossover point}
\label{sec:crossover}

As stated in the main text, the crossover between the scaling $\tau=2$ and
$\tau=3/2$ can be understood as a threshold phenomenon due to the finite temporal
resolution. According to the scaling argument of Eq.~(2) the
crossover is observed when clusters have size large enough for their duration to
be comparable with $\Delta^{-1}$. On average, the largest cluster whose duration
is $T \leq \Delta^{-1}$ has size $S_T \propto \Delta^{-1} T$ so that
$S_T \leq \Delta^{-2}$. Analogously, the smallest cluster whose duration is 
$T \geq \Delta^{-1}$ has, on average, size $S_T \propto T^2$ so that
$S_T \geq \Delta^{-2}$. It follows that the crossover point is expected to
scale with the temporal resolution as $S_c \propto \Delta^{-2}$. This result is
confirmed in Fig.~\ref{fig:crossover}a.
\subsection{System size and sample size}

\begin{figure}[!htb]
\begin{center}
\includegraphics[width=0.95\textwidth]{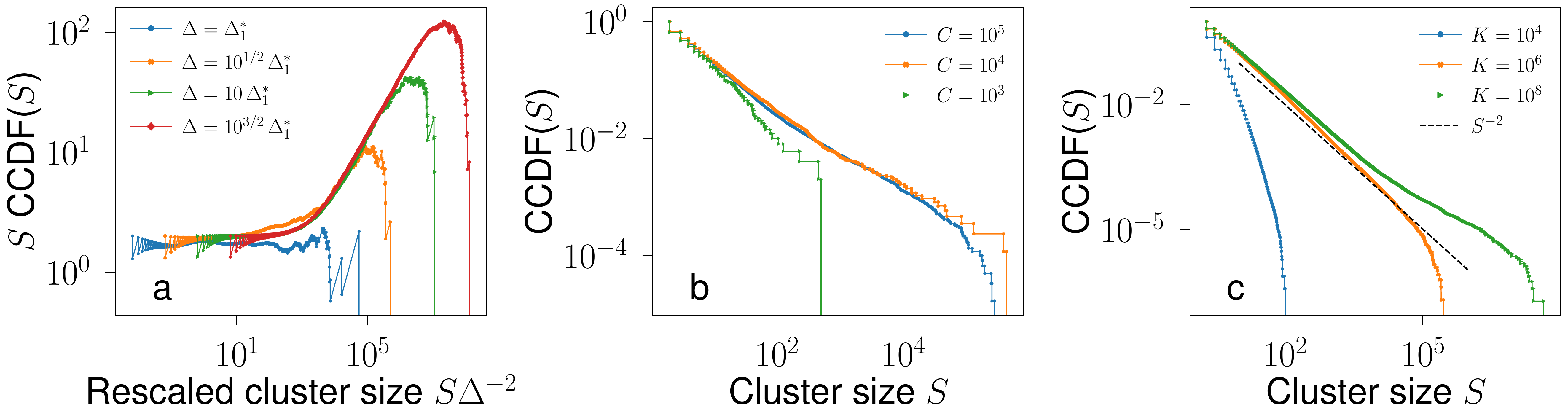}
\end{center}
\caption{Crossover behavior and role of system and sample size in cluster
statistics of the Hawkes process. We use $\phi(x) = e^{-x}$ and $\mu=10^{-4}$. (a) Complementary 
Cumulative Distribution Function (CCDF), rescaled by $S$, against the rescaled
cluster size $S \Delta^{-2}$ of $C=10^7$ clusters. We use here $K=10^8$. (b) CCDF$(S)$ for different sample size $C$. We use here $\Delta=0.1954$ and
$K=10^8$. (c) CCDF$(S)$ for different system size. We consider here $C=10^7$ clusters
and $\Delta=0.0195$. }
\label{fig:crossover}
\end{figure}

The argument further allows us to understand the properties
of the cluster size distribution. Given the power-law shape of $P(S)$, the largest
cluster in a sample of size $C$ 
grows as a power law~\cite{boguna2004cut}.
If $C$ is not large enough for (at least) the largest cluster to have size comparable with
$\Delta^{-1}T$, then the power-law shape of $P(S)$ is not observed at all, and the distribution
appears to be exponential. If the sample size is large enough, then $P(S)$ does not 
show any further dependence on $C$, and the crossover point $S_c$ is not affected
by $C$, as confirmed by Figure~\ref{fig:crossover}b. Finally, for a given system size $K$, we have shown that there
exist a pseudo-critical temporal resolution $\Delta_1^*(K)$ such that $P(S)\sim S^{-2}$. As 
$\Delta_1^* \rightarrow 0$, reducing $K$ for a given $\Delta$ is analogous to increasing $\Delta$ for
a given $K$ (and vice versa). It follows that, for a given temporal resolution $\Delta$, system size can be too small for power-law scaling to be observed or large enough for the crossover between the two regimes to be observed. Between the two extremes, there exists a size such that 
the resolution $\Delta=\Delta_1^*(K)$, as shown in Figure~\ref{fig:crossover}c.

\subsection{Theory of critical branching process explains plateau values observed in phase diagrams}
\label{sec:plateau}

We show here that numerical values of the order parameter and susceptibility 
observed in Figure~1b of the main text, and Figures~\ref{fig:Hawkes_FSS}a, and~\ref{fig:Hawkes_FSS}d can be reproduced by considering the
statistics of the critical branching process. Recall that the order parameter is obtained
as the normalized size of the largest cluster, averaged over several realizations, and
that the susceptibility is defined as the fluctuations of the average size of the 
largest cluster. We can reproduce these statistics by:

\begin{enumerate}

\item Extracting a sample of size $C$ of iid random variables $\{Z_1, Z_2,
  \ldots, Z_C \}$ from the  power-law distribution $P(Z) \sim Z^{-3/2}$, i.e., the distribution
  of the tree size for a critical branching process.

\item Calculating their sum $V = \sum_{c=1}^C Z_q$ and their maximum $Z_{max} = \max \{Z_1, Z_2,
  \ldots, Z_C \}$ to estimate $Y = Z_{max} / V $. In this way $Y$ represents $S_M/K$ in a single
  realization of the Hawkes process. Order parameter and susceptibility can be obtained by averaging.

\item Repeating the first two points $T$ times to obtain the sample
  $\{Y_1, Y_2, \ldots, Y_T\}$, and finally estimating the order
  parameter as
\[
P_\infty = T^{-1} \sum_{t=1}^T Y_t
\]
and the susceptibility as
\[
\chi = \left( T^{-1} \sum_{t=1}^T  Y^2_t - P_\infty^2 \right) / P_\infty
\; .
\]

\end{enumerate}

\begin{figure}[!htb]
\begin{center}
\includegraphics[width=0.95\textwidth]{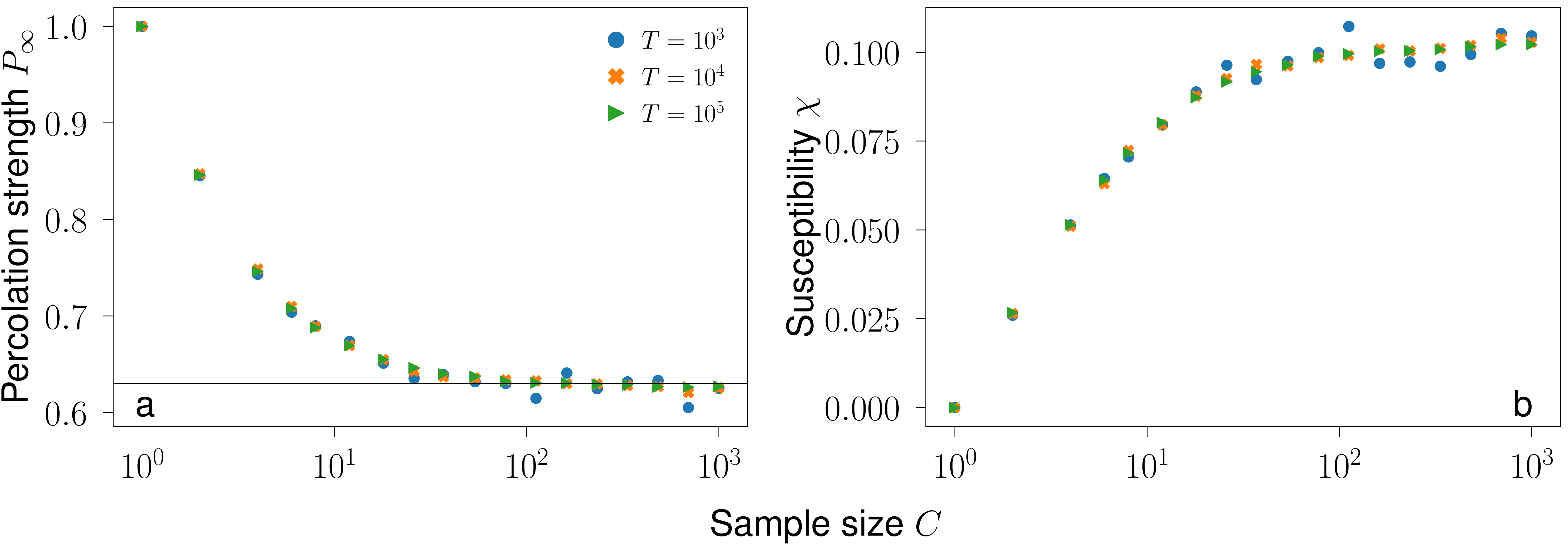}
\end{center}
\caption{(a) Estimate of the order parameter $P_\infty$ as a function 
of the sample size $C$ obtained for different total 
  realizations $T$ of the process. The horizontal black line stands for
  $P_\infty = 1 - 1/e$. 
  (b) Susceptibility $\chi$ as a function of the sample size. $\chi$ plateaus at 
  the same value as observed in Figure~1b of the main text, and Figures~\ref{fig:Hawkes_FSS}a and \ref{fig:Hawkes_FSS}d. }
\end{figure}

\subsection{Percolation theory of the subcritical Hawkes process}
\label{sec:HSUBpercolation}

As stated in the main text, several different systems exhibit 
bursty patterns of activity. Neuronal networks are one possible example
of such systems. Whether the brain truly operates at 
criticality~\cite{beggs2003neuronal} or in a slightly subcritical
state~\cite{priesemann2014spike} is a debated argument.
As such, it could be interesting to check whether the theory presented
in the main text remains valid for $n\lesssim 1$.

\begin{figure}[!htb]
\begin{center}
\includegraphics[width=0.95\textwidth]{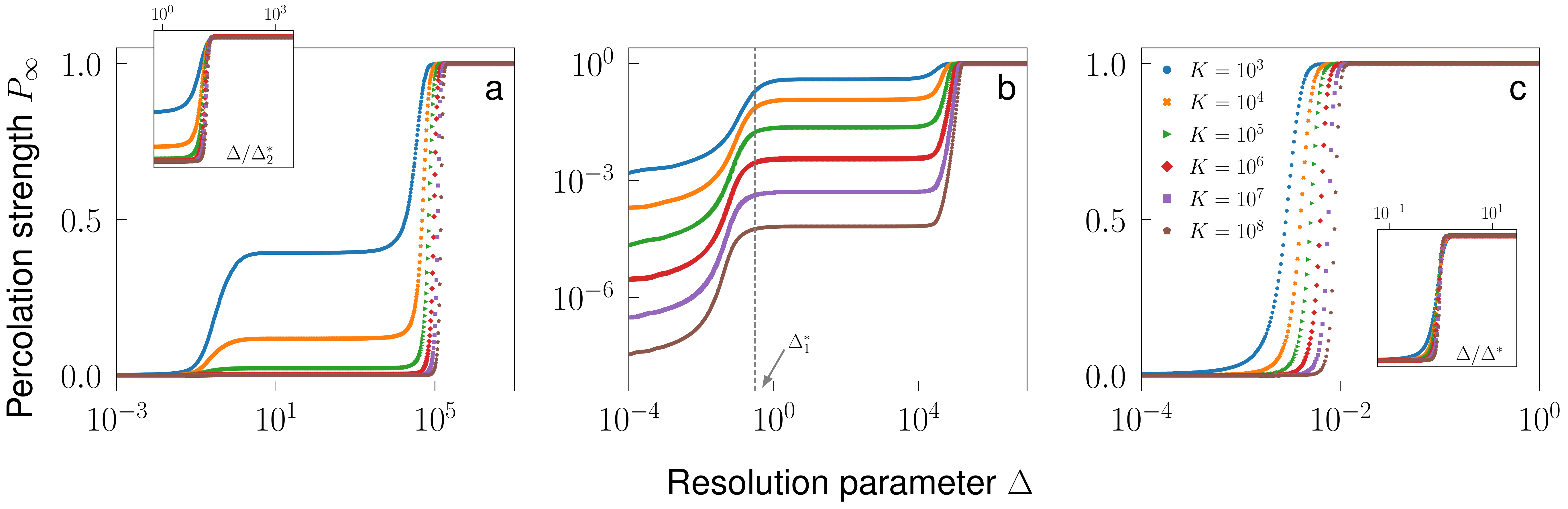}    
\end{center}
\caption{Percolation phase diagrams of the subcritical Hawkes process.
We consider the exponential kernel function and use $n=0.95$. We plot the percolation strength $P_{\infty}$ as a function of the resolution
parameter $\Delta$ for different configurations and different system sizes K. Average
values are obtained by considering $R = 10^3$
realizations of the process. (a) We set $\mu=10^{-4}$. The inset shows the 
same data as of the main panel but the resolution parameter is
rescaled as $\Delta/\Delta^*_2$, with $\Delta^*_2 = (1-n) \log(K) / \mu $. (b) Same
as in panel a, but we use the logarithmic scale for the
ordinates. The vertical gray line roughly indicates the position of
the pseudo-critical point $\Delta^*_1$. We use this value as the
reference for the plots in Figure~\ref{fig:subcrit_distribs}.
(c) We set $\mu=100$. The inset shows the 
same data as of the main panel but the resolution parameter is
rescaled as $\Delta/\Delta^*$, with $\Delta^* = (1-n) \log(K) / \mu $.}
\label{fig:subcrit_PhaseDiag}
\end{figure}

\begin{figure}[!htb]
\begin{center}
\includegraphics[width=0.65\textwidth]{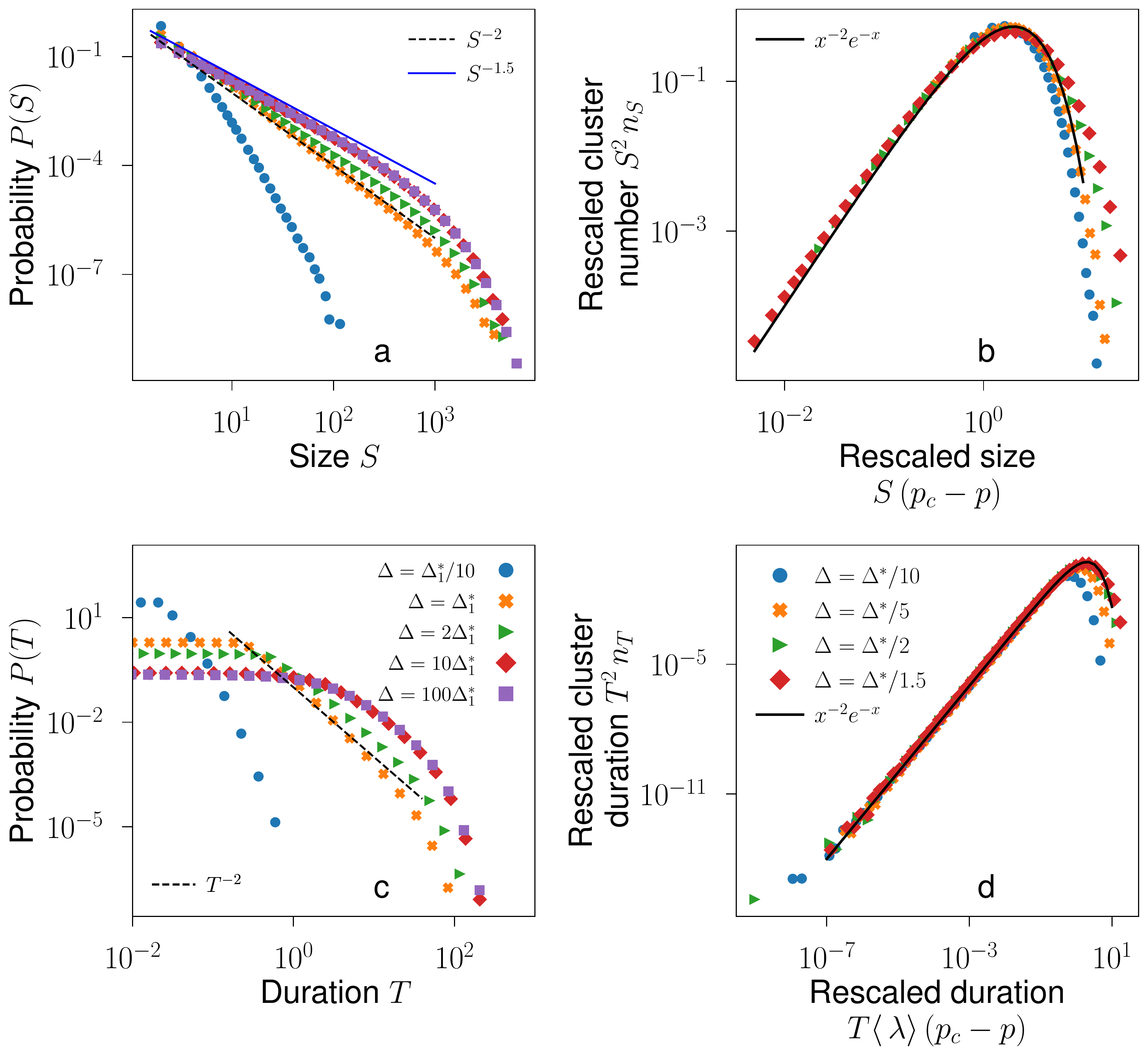}    
\end{center}
\caption{Percolation properties of the subcritical Hawkes
process. We consider temporal processes identical to those studied
in Figure~\ref{fig:subcrit_PhaseDiag}, thus generated using an
exponential kernel function with $n=0.95$. 
System size is fixed at $K = 10^8$. 
Histograms are obtained by considering $C = 10^7$ clusters per configuration. 
(a) We set $\mu=10^{-4}$. Probability density
function of cluster size $S$ for different values of $\Delta$.
The dashed black line scales as $S^{-2}$ while the solid blue line
scales as $S^{-3/2}$.
(b) We set $\mu=100$.
Cluster number $n_s$, rescaled by $S^2$, as a function of
$S(1-p)$, for different values of $p$. The solid black line is the scaling
function $\mathscr{G}(x) = x^2 e^{-x}$. 
(c) We set $\mu=10^{-4}$. Probability density
function of cluster duration $T$ for different values of $\Delta$.
The dashed black line scales as $T^{-2}$.
(d) We set $\mu=100$. Cluster duration $n_T$, rescaled
by $T^2$, as a function of $\av{\lambda} T (1-p)$, for the same values of $p$ as in 
panel (a). The solid black line is the scaling
function $\mathscr{G}(x) = x^2 e^{-x}$. 
}
\label{fig:subcrit_distribs}
\end{figure}

We studied the percolation transition and the cluster distributions in the
Hawkes process with exponential kernel and $n=0.95$.
Results are coherently explained by the theory already used for the critical
configuration. The long-term estimates of the first $\av{\lambda} $ and second moment
$\av{\lambda^2} $ of the rate of an
exponential Hawkes process in the subcritical regime $n<1$ are given again
in Ref.~\cite{dassios2011dynamic}. They read as
\begin{equation}
    \begin{split}
        & \av{\lambda} = \frac{\mu}{1-n} \\
        & \av{\lambda^2} = \frac{2\mu^2 + n^2 \mu}{2(1-n)^2} \, .
    \end{split} 
    \label{eq:av_subcrit_rate}
\end{equation}
Both the above expressions are valid in the long-term limit, when
the process is stationary.
In particular, we note that the square of the ratio standard deviation over average
value is inversely proportional to $\mu$, i.e., 
\begin{equation}
    \frac{\av{\lambda^2} - \av{\lambda}^2}{\av{\lambda}^2} =
    \frac{n^2}{2\mu} \, .
    \label{eq:ratio}
  \end{equation}
  The results of the numerical simulations reported in Figures~\ref{fig:subcrit_PhaseDiag}
and~\ref{fig:subcrit_distribs}  are explained using the above
expressions.

For $\mu \ll 1$, the behavior is similar to the one observed for the
critical Hawkes process. In this regime, time series consist of
quick bursts of activity well separated in time.
The fundamental difference with the critical configuration is that
burst sizes obey a power-law
distribution with a neat exponential cut-off~\cite{kossio2018growing}. As such, the largest cluster 
generated by a single burst has finite size, independently of the
system size $K$. In turn, the value of the plateau observed in the
percolation phase diagram decreases as $K$ increase, and the
associated phase transition disappears in the thermodynamic limit, see Figure~\ref{fig:subcrit_PhaseDiag}.
In finite-size systems, we can still measure the cluster distributions
$P(S)$ and $P(T)$ around the pseudo-critical point $\Delta^*_1$, as
done in  Figure~\ref{fig:subcrit_distribs}. 
The Poisson-like transition is observed at the pseudo-critical point
$\Delta^*_2=\log(K) / \mu$. This second transition does not
vanish as the system size increases.

For $\mu \gg 1$, Eq.~(\ref{eq:ratio}) tells us that fluctuations are
much smaller than the expected value of the rate, so we expect the
process to behave as a Poisson process with rate $\av{\lambda}$.
In particular, the phase diagram
should look like the one of a homogeneous one-dimensional percolation model with 
occupation probability $p=1-e^{-\av{\lambda}\Delta}$, thus consisting of a unique transition
happening at $\Delta^* = \log(K)/\av{\lambda}$. Our prediction
is confirmed by the collapse of
the order parameter shown in Figure~\ref{fig:subcrit_PhaseDiag}c, and
further supported by the numerical analysis regarding the scaling of the cluster number, see
Figure~\ref{fig:subcrit_distribs}.

\changejan{
\subsection{Non-homogeneous Poisson process with rate linearly growing in time}
\label{sec:NHP}

Here we discuss a non-homogeneous Poisson process with rate linearly growing in time, i.e., $\lambda(t) = t$.
Numerical simulations of the model can be performed efficiently by using the inverse transform method. Specifically, the probability that no events are observed in the interval $(t, t+x)$ is 
\begin{equation}
P_{n=0}[t, x] = \exp\left(-\int_t^{t+x} \lambda(t') \, dt' \right) \, .
\end{equation}
Thus, the inter-event time $x$ is a random variable satisfying 
\begin{equation}
    x = -t + \sqrt{t^2 - 2 \log(u)} \, \text{, where } \, u \sim \text{Unif}(0, 1)\, .
    \label{eq:nhp}
\end{equation}
All inter-events are obtained from the above expression, including the first one at $t=0$.

We note that
$\lambda(t) = t$ implies that $k(t) = t^2$ and $\lambda(k) = \sqrt{k}$. 
Figure~\ref{fig:NHP_traject} shows that the expected dependence $k(t) = t^2$ is correct.

\begin{figure}[!htb]
\begin{center}
\includegraphics[width=0.95\textwidth]{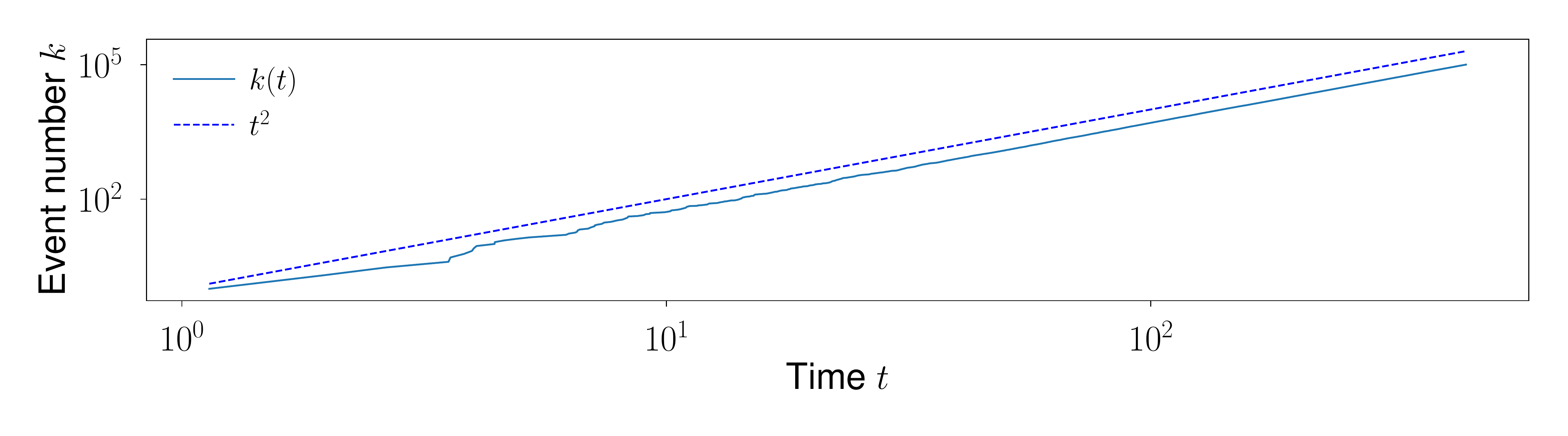}
\end{center}
\caption{Individual realizations of the non-homogeneous Poisson process with deterministic linear time growth. The realization is made of $K=10^5$ events. The dashed blue line is the expected 
relation $k(t) = t^2$.}
\label{fig:NHP_traject}
\end{figure}

We remark that the most general expression for a linearly increasing rate is  $\lambda(t) = \mu + \beta t$. In our experiments, we assume that time is measured in units equal to $\beta$. Further, we assume $\mu =0$. Taking $\mu > 0$ would correspond to a minimal modification of Eq.~(\ref{eq:nhp}) having no significant impact on the results reported below.

Figure~\ref{fig:NHP_clusts}a shows the percolation strength as a function of the resolution parameter 
$\Delta$. The transition point of the model is expected to scale as
$\Delta^* = \log(K)/\sqrt{K}$ for a time series with $K$ events. Such a prediction is verified in the inset of Figure~\ref{fig:NHP_clusts}a.
The validity of the percolation framework is further confirmed in Figure~\ref{fig:NHP_scaling}a,
where the susceptibility is shown to collapse under the usual rescaling of the temporal resolution,
and in Figure~\ref{fig:NHP_scaling}b, where the location of the susceptibility's peak is shown to converge to $0$ as 
$\Delta^*(K) = \log(K)/\sqrt{K}$. Figure~\ref{fig:NHP_scaling}c further displays the measure of the exponent
$\beta=0$ and $\gamma=1$. Figures~\ref{fig:NHP_clusts}b and ~\ref{fig:NHP_clusts}c show the scaling properties of 
finite clusters, revealing that the process belongs to the universality class of 1D
percolation, characterized by the exponents $\tau=\alpha=2$.

\begin{figure}[!htb]
\begin{center}
\includegraphics[width=0.95\textwidth]{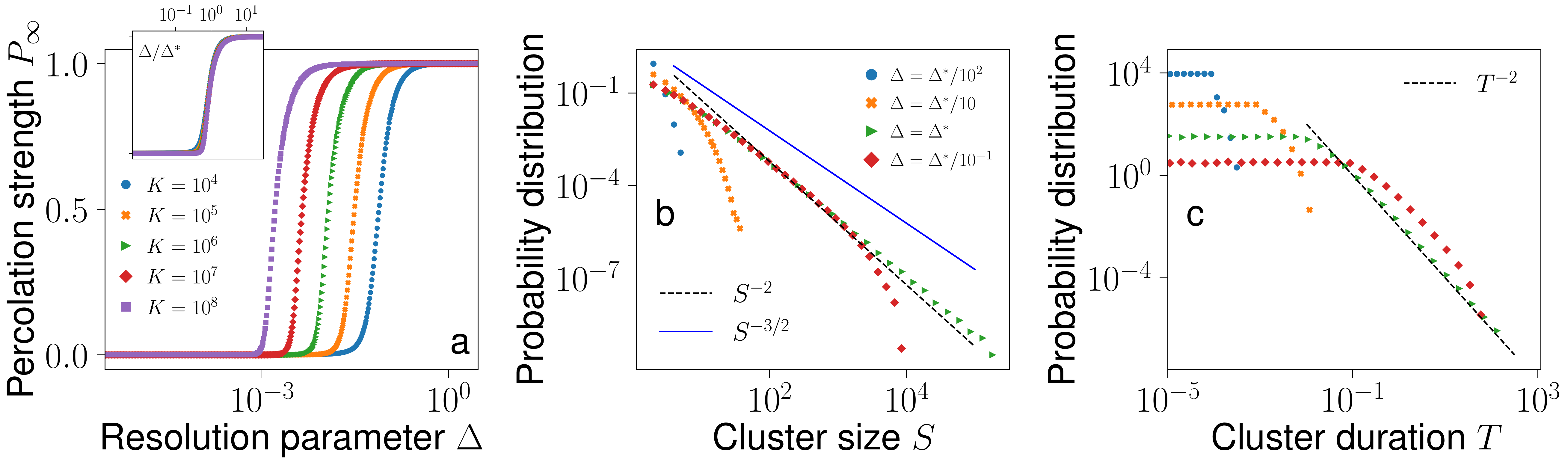}
\end{center}
\caption{Percolation transition in the non-homogeneous Poisson processes. Results are obtained
by considering $R = 10^3$ realizations of the process. (a) Percolation strength as a function of the 
resolution parameter $\Delta$. We display results for
different system sizes $K$. The inset shows the same data as of the main panel 
but the resolution parameter is rescaled as $\Delta/\Delta^*$.
(b) Distribution of the cluster size. System size here is $K = 10^6$. (c) Distribution of the cluster duration for the same data of panel (b).}
\label{fig:NHP_clusts}
\end{figure}

\begin{figure}[!htb]
\begin{center}
\includegraphics[width=0.95\textwidth]{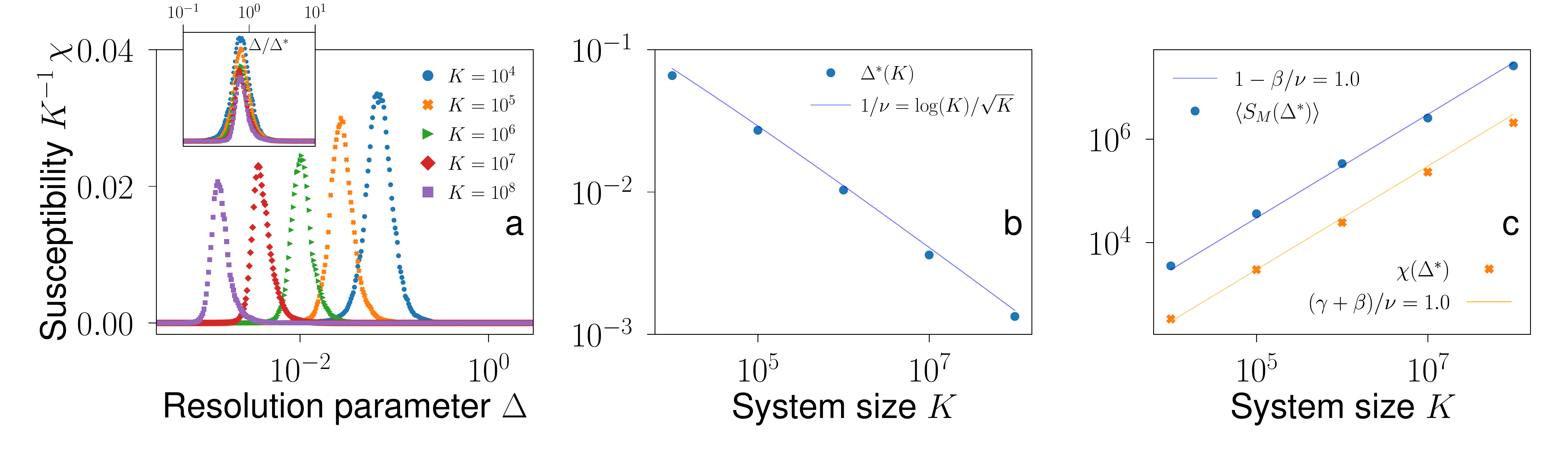}
\end{center}
\caption{Finite-size scaling analysis of percolation in the non-homogeneous Poisson process.(a) Susceptibility [i.e., Eq.~(S5)] divided by K,
as a function of the resolution parameter $\Delta$. The inset shows the same data as of the main panel 
but the resolution parameter is rescaled as $\Delta/\Delta^*$.
(b) Convergence of the effective critical point to the critical point. We measure $\Delta^*$
as the value of the resolution parameter where we observe the peak of the susceptibility. 
(c) Scaling of the peak of susceptibility (orange squares) and of the average size of the largest cluster
(blue circles) at the effective resolution $\Delta^*$.}
\label{fig:NHP_scaling}
\end{figure}

In conclusion, the non-homogeneous Poisson process with linearly increasing rate has the same critical properties as of the homogeneous Poisson process. 
}

\end{document}